\documentclass[10pt,twocolumn,letterpaper]{article}

\usepackage[pagenumbers]{cvpr}

\renewcommand{\paragraph}[1]{\vspace{.5em}\noindent\textbf{#1}}

\usepackage{multirow}
\usepackage{makecell}
\usepackage{tabularx}
\usepackage{microtype}

\providecommand{\valstd}[2]{#1\,{\fontsize{5pt}{6pt}\selectfont$\pm$#2}}
\providecommand{\bvalstd}[2]{\textbf{#1}\,{\fontsize{5pt}{6pt}\selectfont$\pm$#2}}
\providecommand{\uvalstd}[2]{\underline{#1}\,{\fontsize{5pt}{6pt}\selectfont$\pm$#2}}

\usepackage{enumitem}
\setlist[itemize]{leftmargin=*,topsep=4pt,itemsep=3pt,parsep=0pt}
\usepackage{placeins}

\usepackage{needspace}

\usepackage{booktabs}
\usepackage{multirow}
\usepackage{tabularx}
\usepackage{array}

\newcolumntype{Y}{>{\centering\arraybackslash}X}

\providecommand{\gsep}{\rule[-0.25ex]{0.30pt}{1.55ex}}

\definecolor{cvprblue}{rgb}{0.21,0.49,0.74}
\usepackage[breaklinks,colorlinks,allcolors=cvprblue]{hyperref}

\title{TAC: Target-Anchored Coverage for Source--Target Curation in Cross-Center Medical Image Segmentation}
\author{%
Xiwen Chen$^{1}$\quad
Tong Miao$^{1}$\quad
Yueqi Dong$^{1}$\\
Divya Rajparia$^{1, 2}$\quad
Sai Praneeth Karimireddy$^{1}$\quad
Ruishan Liu$^{1}$\\
{\normalsize $^{1}$University of Southern California, USA\quad $^{2}$IIT Hyderabad, India}%
}
\hypersetup{
  pdftitle={TAC: Target-Anchored Coverage for Source-Target Curation in Cross-Center Medical Image Segmentation},
  pdfauthor={Xiwen Chen, Tong Miao, Yueqi Dong, Divya Rajparia, Sai Praneeth Karimireddy, Ruishan Liu}
}

\begin{document}
\maketitle

\begin{abstract}
Cross-center tumor segmentation is constrained by limited target
annotations and heterogeneous labeled source pools whose cases may
transfer poorly to a new center. Target acquisition and source selection
are often treated as separate decisions, leaving acquired target
annotations underused for source curation. We introduce
\textbf{Target-Anchored Coverage (TAC)}, a data-centric framework that
explicitly couples these decisions. TAC acquires an informative, diverse target support set through
weighted clustering guided by foreground-aware uncertainty,
local prediction inconsistency, and prediction stability.
The resulting masks provide not only supervision for final training
but also task-specific anchors for source curation.
TAC estimates target-conditioned source reliability from distributional
similarity, annotation-informed gradient agreement, and
annotation-derived prototype compatibility. It then greedily selects a
complementary source subset by weighting support-capped
facility-location coverage with source reliability.
TAC requires no changes to the segmentation architecture,
supervised loss, or inference procedure. Under leave-one-center-out evaluation on BraTS 2021 brain-tumor MRI
and MAMA-MIA breast-tumor DCE-MRI, TAC achieves
$77.3\%$ macro Dice and $68.4\%$ foreground Dice, respectively,
at representative budgets of $B_t=10$ acquired target masks
and $B_s=150$ selected source cases for final training.
Against the strongest of 20 acquisition--selection combinations
and four domain-adaptation baselines under matched budgets,
TAC gains $1.8$ Dice points on BraTS and $1.9$ on MAMA-MIA.
Budget sweeps show higher mean Dice than the strongest plotted comparator
across tested source budgets and at 8--15 target masks,
with mixed results at smaller target budgets. These findings highlight the dual value of target annotations
for direct supervision and target-conditioned source curation. 
\end{abstract}

\begin{figure}[t]
  \centering
  \includegraphics[width=\columnwidth]{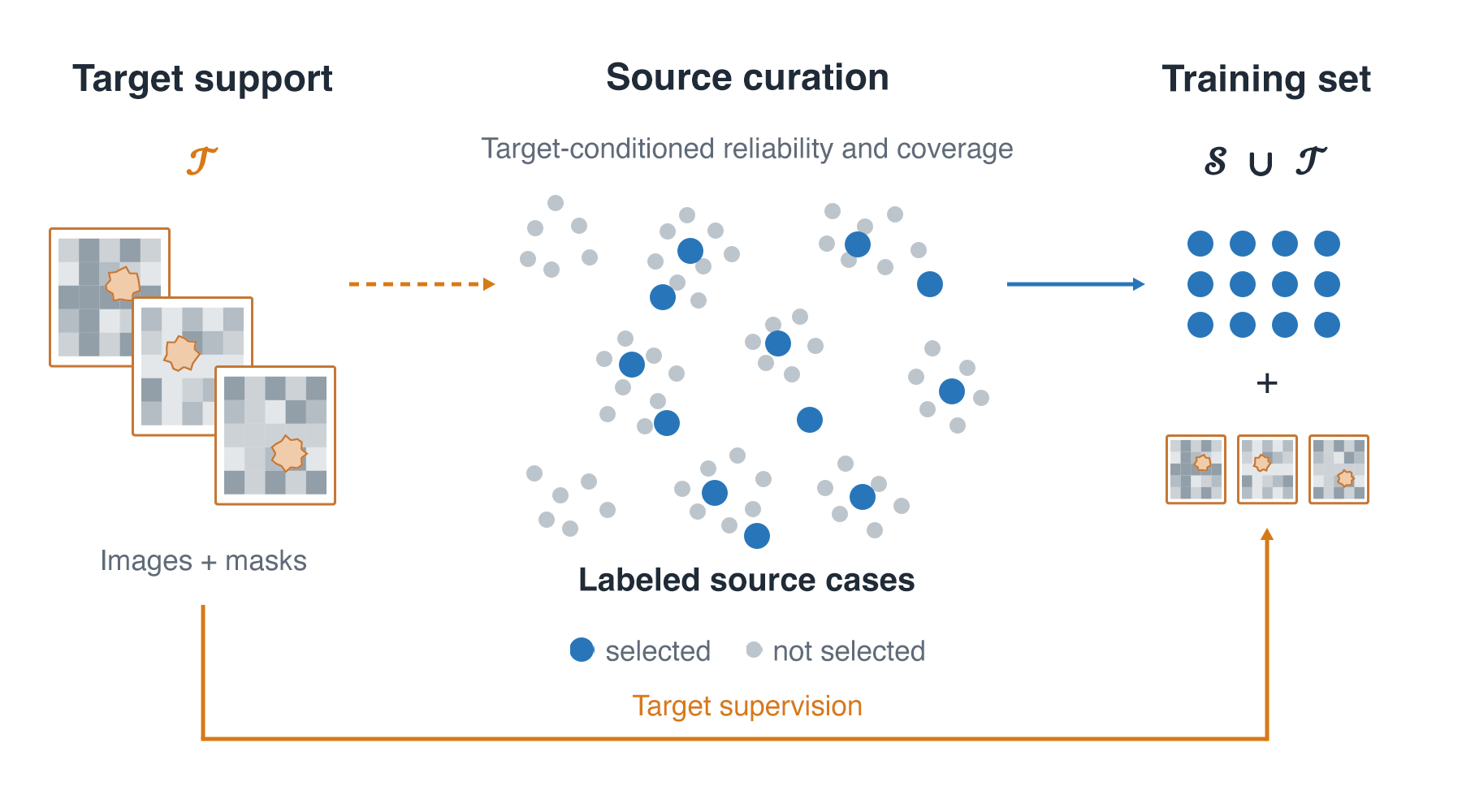}
  \caption{Target support guides source curation and supervises final training.}
  \label{fig:tac_teaser}
\end{figure}

\section{Introduction}
\label{sec:intro}

Generalizing tumor segmentation models to a new imaging center is constrained
by both scarce target annotations and heterogeneous external training data.
Accurate tumor delineation requires expert voxel-level annotation, while
differences in scanners, acquisition protocols, patient populations, and tumor
appearances can substantially affect segmentation performance across centers
\citep{guan2022domainadapt,peng2021limitedsupervision,yoon2024dgreview}.
Although annotated cases from other centers or clinical cohorts may be
available, their utility for a particular target center is not uniform.
Simply pooling all external cases increases the amount of data used for
training and may introduce poorly matched examples that cause negative
transfer~\citep{zhang2023negativetransfer,zhang2023selectivity}.
Effective adaptation therefore depends not only on how a segmentation model
is trained, but also on which target and source cases contribute to its
training set.

We study this problem under two distinct case-level budgets.
Given a heterogeneous labeled source pool and an initially unlabeled
target acquisition pool, a target acquisition budget $B_t$ limits the
number of target cases annotated for training.
A separate source budget $B_s$ limits the number of already-labeled
external cases admitted to final segmentation training.
These constraints give rise to two related decisions:
\emph{which target cases should be annotated, and which external
cases should be selected to complement the acquired target support?}

Existing approaches address different aspects of this problem.
Domain adaptation improves cross-domain transfer through
representation alignment or training-time regularization,
but typically operates on a given source training pool
\citep{guan2022domainadapt,li2023medsegda}.
Data subset selection constructs compact training sets using
representativeness, coverage, gradient-based criteria, or
validation-guided utility
\citep{sener2018coreset,mirzasoleiman2020craig,
killamsetty2021gradmatch,killamsetty2021glister}.
Target-conditioned data selection further exploits target
references to identify valuable training examples
\citep{karanam2022orient,zhang2023selectivity,xia2024less},
but generally treats the target reference as given rather
than deciding which target segmentation masks to acquire.
Conversely, active learning and active domain adaptation
prioritize informative target samples for annotation,
without explicitly using the acquired segmentation masks
to curate a separately budgeted source training subset
\citep{su2020aada,prabhu2021clue,sun2023lada}.
These distinctions motivate a central question:
\emph{how can actively acquired target annotations guide
subsequent source selection under separate annotation
and training budgets?}

Our central insight is that acquired target annotations provide both
supervision and task-specific evidence for source selection. Target images
characterize target-center appearance, whereas expert masks enable gradient-agreement
and prototype-compatibility estimates. Source cases with similar feature-space
relevance may therefore differ in their estimated compatibility with the
target task; these estimates depend on which target cases were acquired.
Ranking sources by individual compatibility alone may yield a redundant
subset concentrated around a narrow range of target-relevant appearances.
Source selection should consider both \emph{reliability}, each candidate's
compatibility with the acquired support, and \emph{coverage}, how well the
subset represents target-relevant source variation. This motivates
\emph{target-first, coupled source--target curation}: target cases are
acquired first, and their annotations guide subsequent source selection.

We introduce \textbf{TAC} (\textbf{T}arget-\textbf{A}nchored
\textbf{C}overage), a data-centric framework that implements this
strategy (\cref{fig:tac_overview}).
TAC first acquires an informative and diverse target support set
by combining foreground-aware uncertainty, local prediction
inconsistency, perturbation stability, and weighted clustering.
Once the selected cases are annotated, TAC estimates the reliability
of each source case using three complementary signals:
distribution match, gradient agreement, and prototype
compatibility.
Their contributions are combined through split-adaptive weights computed
from distribution-match and gradient-agreement stability, alongside
prototype-score spread.
However, high individual reliability alone does not ensure a
well-composed training subset.
TAC therefore uses a support-capped facility-location criterion
to cover support-relevant source regions. Each candidate's
target-conditioned reliability weights its capped similarity inside
the facility-location objective. Additional representatives are rewarded
only where they improve the subset's reliability-weighted coverage. A greedy procedure constructs
the source subset under $B_s$ by selecting the largest marginal gain.
The selected source cases and acquired target annotations are then
used for standard supervised segmentation training, without
changing the segmentation architecture, task loss, or inference
procedure.

We evaluate TAC on BraTS 2021 brain-tumor MRI
\citep{baid2021brats2021} and MAMA-MIA breast-tumor DCE-MRI
\citep{garrucho2025mamamia}, using four held-out target centers
per benchmark.
At the primary setting of $(B_s,B_t)=(150,10)$, TAC achieves
average Dice scores of 77.3\% on BraTS and 68.4\% on MAMA-MIA,
improving over the strongest of 20 budget-matched
target-acquisition and source-selection combinations by
1.8 and 2.0 Dice points, respectively.
TAC also exceeds all four evaluated domain-adaptation baselines
in target-averaged Dice on both benchmarks.
A controlled $2\times2$ acquisition--selection study finds positive
paired-seed interaction terms on both benchmarks.

\noindent\textbf{Our main contributions are as follows:}
\begin{itemize}
    \item We formulate annotation-constrained cross-center tumor
    segmentation as a \emph{coupled source--target curation problem}
    under separate target annotation and source training budgets,
    where acquired target support both supervises training and guides source selection.

    \item We introduce TAC, a target-first curation framework that
    combines split-adaptive, multi-signal source reliability with
    target-anchored facility-location coverage.
    A reliability-weighted facility-location objective selects source
    representatives by their marginal contribution under the source
    budget.

    \item We establish the effectiveness of TAC on two multi-center
    MRI segmentation benchmarks through comparisons with
    20 budget-matched acquisition--selection combinations and
    four domain-adaptation baselines.
    Budget sweeps, component ablations, and controlled
    acquisition--selection experiments further characterize
    its behavior.
\end{itemize}

\section{Related Work}
\label{sec:related}

\paragraph{Domain adaptation for semantic segmentation.}
Domain adaptation mitigates distribution shifts between imaging domains
through feature alignment, image transformation, and training-time
regularization~\citep{guan2022domainadapt,li2023medsegda}.
For semantic segmentation, Fourier Domain Adaptation (FDA) transfers
low-frequency image statistics across domains~\citep{yang2020fda},
whereas Masked Image Consistency (MIC) leverages contextual consistency
on unlabeled target images~\citep{hoyer2023mic}.
When annotations are limited, semi-supervised segmentation can also
exploit unlabeled images through class-wise pixel-level contrastive
learning~\citep{alonso2021pixel}.
For volumetric medical segmentation, MAPSeg explores
unsupervised domain adaptation through 3D masked
autoencoding and pseudo-labeling~\citep{zhang2024mapseg}.
These methods improve cross-domain learning using
available source and target data, whereas TAC additionally
considers how to construct the training set under separate
annotation and source-selection budgets.

\paragraph{Source subset selection and target-conditioned valuation.}
Data-efficient training selects representative cases using coresets
or submodular objectives~\citep{sener2018coreset,mirzasoleiman2020craig,wei2015submodularity,guo2022deepcore},
or preserves aggregate training signals through gradient
matching~\citep{killamsetty2021gradmatch}.
Selection via Proxy uses inexpensive proxy models to reduce
data-selection cost~\citep{coleman2020selection}.
Data pruning also examines how selection quality affects
learning curves~\citep{sorscher2022pruning}.
Data valuation estimates individual contributions through influence
functions, Shapley values, or training trajectories
\citep{koh2017influence,ghorbani2019datashapley,pruthi2020tracin}.
Optimization-aware selection further evaluates training examples
against a reference objective: GLISTER optimizes held-out validation
performance~\citep{killamsetty2021glister}, while LESS estimates
target-task influence using optimizer-aware gradient representations
for instruction-tuning data selection~\citep{xia2024less}.
Transfer-oriented dataset pruning also studies source-class
selection based on source--target label or feature
mappings~\citep{zhang2023selectivity}.
Target-guided submodular methods explicitly account for both
reference relevance and set composition: PRISM develops submodular
information measures for relevance, diversity, and
coverage~\citep{kothawade2022prism}, while ORIENT selects source cases
relevant to a labeled target reference using submodular mutual information
on gradient similarities~\citep{karanam2022orient}.
These target-conditioned approaches demonstrate the value of a
reference-guided source subset, but typically assume that the target
reference is already available. Constructing that reference is itself
a data acquisition problem.

\paragraph{Active target acquisition under domain shift.}
Active learning reduces annotation demand by querying uncertain,
representative, or diverse examples
\citep{settles2009active,gal2017bayesianactive,sener2018coreset,ash2020badge}.
BatchBALD selects batches using their joint information about model
parameters~\citep{kirsch2019batchbald}.
Submodular active learning further considers challenging
candidate pools containing redundant samples, rare classes,
or out-of-distribution examples~\citep{kothawade2021similar}.
Under domain shift, AADA and transferable query selection consider
domain relevance alongside annotation
informativeness~\citep{su2020aada,fu2021tqs}.
CLUE clusters uncertainty-weighted embeddings~\citep{prabhu2021clue},
LADA exploits local prediction context~\citep{sun2023lada},
and MADA combines active target querying with multi-anchor
adaptation~\citep{ning2021mada}.
Medical active learning addresses expert annotation costs
\citep{budd2021survey,wang2024medicalal} through uncertainty and
image similarity~\citep{yang2017suggestiveannotation,nath2021active},
stochastic batches~\citep{gaillochet2023stochastic},
uncertainty-aware submodular mutual information~\citep{follmer2024usim},
and metric-learning coresets for 3D segmentation~\citep{vepa2024dmlcoreset}.
TAC additionally uses acquired masks to curate labeled source data.

\paragraph{Joint source--target data curation.}
Prior work has connected source and target data decisions
through joint source reweighting and target querying~\citep{chattopadhyay2013joint},
while Bi3D selects target-like source examples before
querying target annotations~\citep{yuan2023bi3d}.
TAC considers a complementary setting with separate
case-level budgets for target annotation ($B_t$) and
source training ($B_s$).
Its central insight is to transform actively acquired
target masks into task-specific evidence for source curation.
Beyond image-level similarity, these annotations enable
gradient-agreement and prototype-compatibility estimates that
inform source reliability.
TAC combines this evidence with target-anchored coverage
to select a reliable yet complementary source subset.

\section{Method}
\label{sec:method}

\begin{figure*}[!t]
  \centering
  \includegraphics[width=\textwidth]{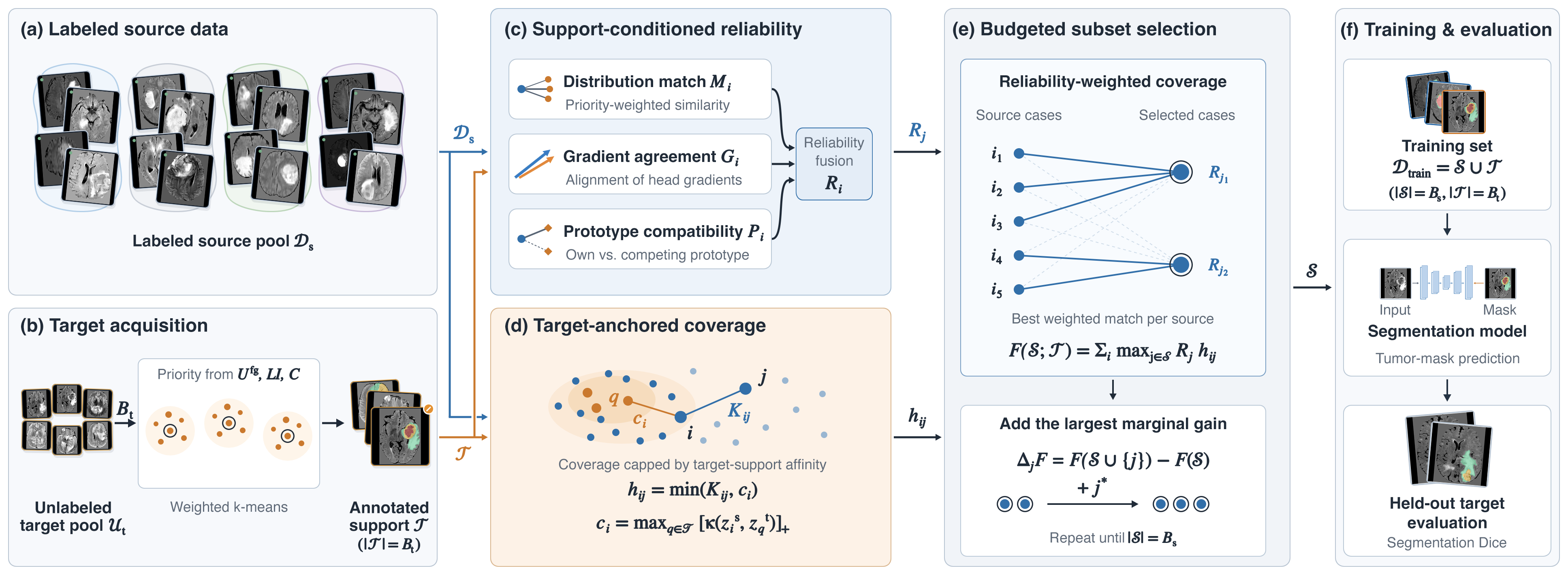}
  \caption{\textbf{TAC overview.} (a,b) Labeled sources and acquired target support inform (c) source reliability and (d) target-anchored coverage via PRISM's FLVMI~\citep{kothawade2022prism}. (e) Greedy selection uses reliability-weighted coverage gains under the source budget. (f) Selected sources and target support train the segmenter. Real BraTS slices; overlays: reference masks.}
  \label{fig:tac_overview}
\end{figure*}

\subsection{Problem Formulation and Overview}
\label{subsec:problem_design}
We study annotation-constrained cross-center segmentation under a
leave-one-center-out protocol. Let
\(\mathcal{D}_s=\{(x_i^s,y_i^s)\}_{i=1}^{N_s}\) denote the labeled source pool
from the remaining centers and
\(\mathcal{U}_t=\{x_j^t\}_{j=1}^{N_t}\) the unlabeled candidate pool from the
held-out target center's training split. Given a target acquisition budget \(B_t\) and a
final-training source budget \(B_s\), TAC acquires and annotates \(B_t\) candidates to form
the support \(\mathcal{T}\) and selects \(\mathcal{S}\subseteq\mathcal{D}_s\)
with \(|\mathcal{S}|=B_s\). The segmenter is trained on
\(\mathcal{S}\cup\mathcal{T}\) without modifying its architecture, supervised
loss, or inference procedure. The source-trained warm-up accesses the full
labeled source pool; \(B_s\) limits only final training. Target validation
annotations are additional to \(B_t\) and used only for checkpoint selection.

The acquired support supervises final training and conditions source selection. TAC comprises target acquisition, source reliability
estimation, target-anchored coverage, and reliability-weighted source
selection (\cref{fig:tac_overview}). Selection runs once before final training using frozen warm-up evidence
and acquired target annotations
(\cref*{app:selection_time_protocol}). Throughout, \(\kappa\) denotes cosine
similarity, \([u]_+=\max(u,0)\), and \(\operatorname{Norm}\) min--max
normalization over the cases being scored (\cref*{app:normalization}).

\subsection{Selection-Time State and Target Acquisition}
\label{subsec:target_anchor}
\paragraph{Selection-time representations.}
The frozen final warm-up checkpoint processes all slices of source cases
and target candidates. For each case, pooled backbone features are
averaged over the top \(30\%\) of slices ranked by a combined
foreground--entropy score. The mean is \(\ell_2\)-normalized to obtain
\(z_i^s\) for source case \(i\) and \(z_j^t\) for target candidate \(j\).
The same target slices provide the acquisition statistics.

\paragraph{Target acquisition.}
Inspired by uncertainty-weighted clustering in CLUE~\citep{prabhu2021clue}
and local inconsistency in LADA~\citep{sun2023lada}, TAC assigns each
candidate \(x_j^t\in\mathcal{U}_t\) the priority
\[
a_j=\operatorname{Norm}(U_j^{\mathrm{fg}})
+0.5\operatorname{Norm}(LI_j)+0.15\operatorname{Norm}(C_j),
\]
where \(U_j^{\mathrm{fg}}\) is mean entropy multiplied by mean foreground
probability, \(LI_j\) aggregates expected label mismatch across neighboring
mean prediction profiles, and \(C_j\) measures prediction agreement under
intensity perturbations. The positive weight on \(C_j\) favors candidates with stable predictions. Weighted \(k\)-means partitions the candidate
embeddings into \(B_t\) clusters with sample weights \(\max(a_j,10^{-6})\). Each cluster
contributes one case balancing centroid proximity and priority
(\cref*{app:target_acquisition_details}); the selected cases are annotated
to form the target support set \(\mathcal{T}\).

\subsection{Target-Conditioned Source Reliability}
\label{subsec:source_reliability}
The acquired support provides image-level reference information for
distribution matching and annotation-derived evidence for gradient
agreement and prototype compatibility.

\paragraph{Distribution match.}
Let \(\mathcal{N}(i)\) contain the \(\min(8,B_t)\) support cases with the
largest clipped cosine similarity to \(z_i^s\). With acquisition priorities divided by their stabilized
neighborhood sum to give \(\bar a_{ir}\), define
\[
M_i=\operatorname{Norm}\!\left(
\sum_{r\in\mathcal{N}(i)}\bar a_{ir}
[\kappa(z_i^s,z_r^t)]_+\right).
\]\paragraph{Gradient agreement.}
At the final warm-up epoch $e_w$ (5 for BraTS; 25 for MAMA-MIA), let \(g_i^{(e_w)}\) denote the
\(\ell_2\)-normalized loss gradient of source case \(i\) with respect to
the final segmentation-head parameters, and \(\bar g^{(e_w)}\) the mean of the
normalized support-case gradients. Define
\[
G_i=\operatorname{Norm}\!\left([\kappa(g_i^{(e_w)},\bar g^{(e_w)})]_+\right).
\]

\paragraph{Prototype compatibility.}
Support cases are grouped into up to three annotation-derived classes using
support tertiles of a tumor-composition score for BraTS or tumor fraction for
MAMA-MIA. Let \(\pi_c\) be the normalized mean embedding of the support cases
in nonempty class \(c\). Source case \(i\) is assigned class \(c(i)\) using the
same dataset-specific scoring rule, support-derived feature scaling for BraTS,
and support-derived thresholds. Compatibility contrasts similarity to the
assigned prototype with that to the most similar prototype from another nonempty class:
\[
P_i=\operatorname{Norm}\!\left(
[\kappa(z_i^s,\pi_{c(i)})]_+
-\max_{c\neq c(i)}[\kappa(z_i^s,\pi_c)]_+\right).
\]
Class scores and empty- or single-class rules appear in
\cref*{app:reliability_defaults}.

\paragraph{Reliability fusion.}
Split-adaptive weights use resampling stability for $M$ and $G$
and score spread for $P$:
\[
\begin{aligned}
R_i &= \lambda_M M_i+\lambda_G G_i+\lambda_P P_i,\\
\lambda_k &=
\frac{\exp(s_k/\tau)}
{\sum_{\ell\in\{M,G,P\}}\exp(s_\ell/\tau)}.
\end{aligned}
\]
For $X\in\{M,G\}$, let $X^{(v)}$ denote scores recomputed on each of
$V=8$ uniformly sampled support subsets of size
$\max(2,\operatorname{round}(0.8B_t))$, drawn without replacement. Define
\[
s_X=\frac{1}{2V}\sum_{v=1}^{V}
\left[\frac{\rho(X,X^{(v)})+1}{2}+o_n(X,X^{(v)})\right],
\]
where $\rho$ is Pearson correlation over source cases and $o_n$ is the
fraction of shared top-$n$ cases, with $n=\min(150,N_s)$.
For $B_t\le2$, $s_M=1$ instead.
We set $s_P=0.5+0.5\,\operatorname{std}(P)$ using the population standard
deviation over source cases, and $\tau=0.3$. The resulting convex combination
lies in $[0,1]$ and requires no further normalization.
Resampling, degenerate-correlation, and tie rules appear in
\cref*{app:reliability_defaults}.

\subsection{Target-Anchored Coverage}
\label{subsec:coverage_objective}
Coverage complements individual reliability by encouraging representation
of target-relevant source regions. Let \(K_{ij}=[\kappa(z_i^s,z_j^s)]_+\) denote source similarity and
\(c_i=\max_{q\in\mathcal{T}}[\kappa(z_i^s,z_q^t)]_+\) the affinity of
source case \(i\) to the support. Motivated by target-guided submodular
selection~\citep{kothawade2022prism,kothawade2022talisman,karanam2022orient},
we define the support-capped facility-location criterion, with
\(h_{ij}=\min(K_{ij},c_i)\), as
\[
C(\mathcal{S};\mathcal{T})=\sum_{i\in\mathcal{D}_s}
\max_{j\in\mathcal{S}}h_{ij}.
\]
The construction follows the FLVMI form in PRISM with relevance scaling fixed to one
(\cref*{app:unified_selector}). The cap \(c_i\) limits the coverage contribution of
source case \(i\) by its affinity to the acquired support. The facility-location maximum rewards
additional representatives only where they improve this capped coverage. The maximum over an empty set is defined as zero.

\subsection{Reliability-Weighted Source Selection}
\label{subsec:adaptive_source_selection}
Representing source case \(i\) by a selected case \(j\) yields utility
\(R_jh_{ij}\), and under \(|\mathcal{S}|=B_s\) we maximize
\[
F(\mathcal{S};\mathcal{T})=\sum_{i\in\mathcal{D}_s}
\max_{j\in\mathcal{S}}R_jh_{ij}.
\]
Reliability \(R_j\) scales the coverage provided by representative \(j\);
setting \(R_j=1\) recovers \(C\).

\paragraph{Greedy selection.}
Let \(m_i(\mathcal{S})=\max_{k\in\mathcal{S}}R_kh_{ik}\), with
\(m_i(\emptyset)=0\). Starting from \(\mathcal{S}=\emptyset\), each step adds
the unselected case \(j^\star\) maximizing
\[
\Delta_jF=\sum_{i\in\mathcal{D}_s}\big[R_jh_{ij}-m_i(\mathcal{S})\big]_+
\]
and sets \(m_i\leftarrow\max(m_i,R_{j^\star}h_{ij^\star})\) for all \(i\),
until \(|\mathcal{S}|=B_s\). Ties are resolved by larger \(R_j\), then fixed source-case order,
also when all gains are zero. For fixed nonnegative \(R_j\) and \(h_{ij}\), each
facility-location term is nondecreasing with diminishing returns. Thus,
\(F\) is monotone and submodular with \(F(\emptyset)=0\). Greedy
selection therefore achieves a \((1-1/e)\) approximation for \(F\) under the cardinality
budget~\citep{nemhauser1978analysis}.

\section{Experiments}
\label{sec:experiments}

\subsection{Experimental Setup}
\label{subsec:exp_setup}

\paragraph{Benchmarks and evaluation.}
We evaluate TAC on BraTS 2021 brain-tumor MRI~\citep{baid2021brats2021}
and MAMA-MIA breast DCE-MRI~\citep{garrucho2025mamamia}.
Our BraTS experiments use 1,251 labeled four-sequence MRI cases
from 15 collection-defined domains, while MAMA-MIA contains
1,506 cases from four cohorts. We treat these domains and cohorts
as centers in our cross-center evaluation. We evaluate four held-out targets per benchmark:
C4, C5, TCGA-LGG, and TCGA-GBM for BraTS; NACT, I-SPY1, Duke,
and I-SPY2 for MAMA-MIA. For each target, all other centers
form the labeled source pool.

For each target, we reserve a fixed, patient-disjoint test set
and approximately 15\% of all cases (at least five) for validation.
The remaining non-test cases form the initially unlabeled
acquisition pool. The same splits are used across methods,
budgets, and seeds. Validation annotations are excluded from $B_t$
and used only for checkpoint selection; test labels are used only
for final evaluation. Target split sizes appear in
\cref{tab:tac_split_summary}; full budget fractions and split details
are given in Supplementary~\cref*{app:split_details}.

\paragraph{Budgets and metrics.}
We sweep source budgets $B_s\in\{50,100,150,200\}$ with $B_t=10$
and target annotation budgets $B_t\in\{2,5,8,10,15\}$
with $B_s=150$.
The primary setting is $(B_s,B_t)=(150,10)$. We report 3D case-level
Macro Dice over ET, TC, and WT for BraTS and foreground Dice for
MAMA-MIA. Dataset-level averages weight the four targets equally.

\begin{table}[t]
    \centering
    \caption{\textbf{Target splits and annotation fractions.} Cand.: unlabeled acquisition pool; Val./Test: held out; Src.: labeled cases from other centers. Last column: $100B_t/\mathrm{Cand.}$ at $B_t=10$. All budgets: Supplementary~\cref*{tab:tac_protocol_splits}.}
    \label{tab:tac_split_summary}
    \footnotesize
    \setlength{\tabcolsep}{2.4pt}
    \renewcommand{\arraystretch}{1.06}
    \begin{tabular}{@{}llrrrrr@{}}
        \toprule
        Dataset & Target & Cand. & Val. & Test & Src. & $B_t=10$ (\%) \\
        \midrule
        \multirow{4}{*}{BraTS}
            & C4  & 15 & 5  & 10 & 1221 & 66.7 \\
            & C5  & 26 & 7  & 14 & 1204 & 38.5 \\
            & LGG & 35 & 10 & 20 & 1186 & 28.6 \\
            & GBM & 55 & 15 & 32 & 1149 & 18.2 \\
        \midrule
        \multirow{4}{*}{MAMA-MIA}
            & NACT   & 19  & 10  & 35  & 1442 & 52.6 \\
            & I-SPY1 & 40  & 26  & 105 & 1335 & 25.0 \\
            & Duke   & 63  & 44  & 184 & 1215 & 15.9 \\
            & I-SPY2 & 172 & 147 & 661 & 526  & 5.8 \\
        \bottomrule
    \end{tabular}
\end{table}

\paragraph{Baselines.}
At the primary budget, we compare TAC against 20 combinations
formed by crossing five target acquisition policies (Random,
CoreSet~\citep{sener2018coreset}, AADA~\citep{su2020aada},
CLUE~\citep{prabhu2021clue}, and LADA~\citep{sun2023lada})
with four source selectors (Random, CoreSet~\citep{sener2018coreset},
LESS~\citep{xia2024less}, and ORIENT~\citep{karanam2022orient}).
Budget sweeps use the 15 combinations excluding LESS.
Four domain-adaptation baselines comprise FDA~\citep{yang2020fda} and
MIC~\citep{hoyer2023mic} (UDA methods, here additionally supervised with
$B_t$ randomly selected labeled target training cases), pixel-level contrastive
SSDA~\citep{alonso2021pixel}, and MADA~\citep{ning2021mada} (ADA).
All budgeted methods use the corresponding labeled-data budgets at
each sweep point. Full Source-only,
Random + Full Source, and Random target-only serve as non-budget-matched
references. Detailed baseline and adaptation protocols are given in
Supplementary~\cref*{app:baseline_protocol,app:da_protocol,app:budget_details}.

\paragraph{Implementation and reporting.}
BraTS uses four-channel EfficientViT-Seg-L1 initialized from
ADE20K-pretrained weights~\citep{cai2023efficientvit,zhou2017scene}
and trained for 20 epochs. MAMA-MIA uses a three-channel 2D
nnU-Net v2~\citep{isensee2021nnunet}, initialized from scratch
and trained for 100 epochs. Within each benchmark, TAC and the
selection-only combinations share the supervised architecture,
loss, and training schedule; adaptation methods use the same
benchmark-specific backbone and supervised loss with their own
adaptation objectives. Selection evidence is extracted from the final checkpoint
of a full-source warm-up: 5 epochs for BraTS and 25 epochs
for MAMA-MIA, corresponding to one-quarter of each
training schedule. The budget $B_s$ limits the source cases
used in final segmentation training.

We repeat the acquisition, source selection, and final training
pipeline with three independent seeds, using the same predefined
validation-based checkpoint-selection rule.
Per-center results report mean $\pm$ sample standard deviation;
Avg. is the mean of the three seed-level four-center averages. Additional training, selection-time,
and reporting details are given in
Supplementary~\cref*{app:brats_training,app:mamamia_training,app:selection_time_protocol,app:reporting_protocol}.

\subsection{Main Results under Matched Budgets}
\label{subsec:main_results}

\begin{table*}[t]
\centering
\caption{\textbf{Per-target Dice (\%)}, $(B_s,B_t)=(150,10)$. LGG/GBM: TCGA-LGG/TCGA-GBM. Entries: mean $\pm$ sample std over three independent seeds; Avg.: unweighted four-center mean. Full Source-only uses no target labels. Reference configurations are excluded from budget-matched rankings. Bold/underlined: best/second-best means, ranked before rounding.}
\label{tab:tac_main_results}
\fontsize{6.5pt}{7.4pt}\selectfont
\setlength{\tabcolsep}{0.9pt}
\renewcommand{\arraystretch}{1.04}

\begin{tabularx}{\textwidth}{@{}l@{\hspace{3pt}\gsep\hspace{3pt}}l*{5}{Y}@{\hspace{3pt}}c@{\hspace{3pt}}*{5}{Y}@{}}
\toprule
\multirow{2}{*}{\shortstack[l]{Target\\acquisition}}
& \multirow{2}{*}{\shortstack[l]{Source\\selection}}
& \multicolumn{5}{c}{BraTS 2021 Macro Dice $\uparrow$}
& \gsep
& \multicolumn{5}{c}{MAMA-MIA Foreground Dice $\uparrow$} \\
\cmidrule(lr){3-7}\cmidrule(lr){9-13}
& & C4 & C5 & LGG & GBM & Avg.
& \gsep
& NACT & I-SPY1 & Duke & I-SPY2 & Avg. \\
\midrule
\multicolumn{2}{l}{Full Source-only} & \valstd{73.2}{0.3} & \valstd{76.7}{1.6} & \valstd{61.0}{1.5} & \valstd{83.7}{0.8} & 73.6 & \gsep & \valstd{67.0}{1.1} & \valstd{67.5}{2.8} & \valstd{55.4}{2.1} & \valstd{67.7}{0.8} & 64.4 \\
\multicolumn{2}{l}{Random + Full Source} & \valstd{75.0}{0.8} & \valstd{77.3}{1.4} & \valstd{62.5}{0.9} & \valstd{83.9}{0.2} & 74.7 & \gsep & \valstd{67.8}{1.8} & \valstd{69.9}{1.1} & \valstd{58.2}{2.5} & \valstd{69.1}{1.8} & 66.3 \\
\multicolumn{2}{l}{Random target-only} & \valstd{68.8}{1.3} & \valstd{72.6}{0.9} & \valstd{50.6}{1.0} & \valstd{77.9}{2.1} & 67.5 & \gsep & \valstd{49.4}{1.2} & \valstd{57.9}{1.5} & \valstd{52.2}{1.2} & \valstd{62.4}{0.8} & 55.5 \\
\midrule
\multirow{4}{*}{Random} & Random & \valstd{73.9}{1.3} & \valstd{77.6}{0.4} & \valstd{60.0}{2.1} & \valstd{82.9}{0.2} & 73.6 & \gsep & \valstd{65.2}{0.2} & \valstd{66.3}{0.4} & \valstd{56.2}{1.7} & \valstd{66.6}{1.4} & 63.6 \\
 & CoreSet~\citep{sener2018coreset} & \valstd{74.3}{0.1} & \valstd{77.6}{1.1} & \valstd{62.1}{1.0} & \valstd{82.8}{0.6} & 74.2 & \gsep & \valstd{67.2}{0.5} & \valstd{67.4}{0.1} & \valstd{56.4}{0.5} & \valstd{65.7}{0.2} & 64.2 \\
 & LESS~\citep{xia2024less} & \valstd{73.1}{0.6} & \valstd{76.7}{0.3} & \valstd{61.7}{3.9} & \valstd{82.6}{0.4} & 73.5 & \gsep & \valstd{60.7}{0.8} & \valstd{69.0}{0.1} & \uvalstd{61.4}{0.4} & \valstd{65.9}{0.4} & 64.3 \\
 & ORIENT~\citep{karanam2022orient} & \valstd{75.4}{0.6} & \valstd{77.6}{0.2} & \valstd{63.7}{1.7} & \bvalstd{83.7}{0.3} & 75.1 & \gsep & \valstd{62.2}{1.4} & \valstd{67.0}{0.1} & \valstd{55.3}{0.4} & \valstd{66.2}{0.5} & 62.7 \\
\midrule
\multirow{4}{*}{CoreSet~\citep{sener2018coreset}} & Random & \valstd{74.9}{0.1} & \valstd{77.9}{0.2} & \valstd{58.7}{2.6} & \valstd{82.9}{0.1} & 73.6 & \gsep & \valstd{68.0}{0.3} & \valstd{68.0}{0.5} & \valstd{59.8}{0.4} & \uvalstd{69.7}{0.2} & 66.4 \\
 & CoreSet & \valstd{75.3}{0.3} & \valstd{78.3}{0.7} & \valstd{61.5}{0.2} & \valstd{82.9}{0.4} & 74.5 & \gsep & \valstd{68.0}{0.5} & \valstd{66.1}{0.3} & \valstd{55.2}{0.6} & \valstd{66.2}{0.3} & 63.9 \\
 & LESS & \valstd{74.1}{0.1} & \valstd{77.8}{0.3} & \valstd{61.5}{0.2} & \valstd{82.5}{0.3} & 74.0 & \gsep & \valstd{61.5}{0.4} & \valstd{67.3}{0.3} & \valstd{58.6}{0.4} & \valstd{66.0}{0.4} & 63.3 \\
 & ORIENT & \uvalstd{77.8}{0.9} & \valstd{78.1}{0.6} & \valstd{61.3}{1.0} & \valstd{82.6}{0.5} & 75.0 & \gsep & \valstd{67.0}{1.0} & \valstd{65.3}{0.4} & \valstd{54.1}{0.2} & \valstd{67.9}{0.1} & 63.6 \\
\midrule
\multirow{4}{*}{AADA~\citep{su2020aada}} & Random & \valstd{76.4}{1.6} & \valstd{77.0}{0.3} & \valstd{59.8}{1.3} & \valstd{82.8}{0.1} & 74.0 & \gsep & \valstd{64.1}{1.3} & \valstd{68.6}{0.2} & \valstd{58.4}{0.2} & \valstd{68.6}{0.9} & 64.9 \\
 & CoreSet & \valstd{75.7}{0.1} & \valstd{77.8}{0.9} & \valstd{63.9}{1.8} & \valstd{82.8}{0.2} & 75.1 & \gsep & \valstd{63.7}{1.7} & \valstd{67.1}{0.1} & \valstd{50.9}{2.7} & \valstd{66.9}{0.2} & 62.1 \\
 & LESS & \valstd{75.9}{1.6} & \valstd{76.4}{1.1} & \valstd{62.6}{0.7} & \valstd{82.3}{0.1} & 74.3 & \gsep & \valstd{62.9}{0.3} & \valstd{69.0}{0.2} & \valstd{59.2}{0.5} & \valstd{65.7}{0.2} & 64.2 \\
 & ORIENT & \valstd{77.4}{2.9} & \uvalstd{78.4}{0.5} & \valstd{63.0}{1.9} & \valstd{83.0}{0.2} & 75.5 & \gsep & \valstd{67.3}{0.1} & \valstd{69.3}{0.3} & \valstd{55.8}{0.1} & \valstd{67.8}{0.6} & 65.0 \\
\midrule
\multirow{4}{*}{CLUE~\citep{prabhu2021clue}} & Random & \valstd{74.5}{0.8} & \valstd{76.9}{0.4} & \valstd{61.6}{1.7} & \valstd{83.1}{0.1} & 74.0 & \gsep & \valstd{65.1}{0.5} & \valstd{68.3}{0.1} & \valstd{59.0}{0.6} & \valstd{69.5}{0.4} & 65.5 \\
 & CoreSet & \valstd{75.6}{1.0} & \valstd{77.8}{0.1} & \valstd{63.0}{0.7} & \valstd{82.9}{0.1} & 74.8 & \gsep & \valstd{66.0}{0.3} & \valstd{67.8}{0.1} & \valstd{52.8}{2.5} & \valstd{66.9}{0.1} & 63.4 \\
 & LESS & \valstd{74.3}{0.4} & \valstd{77.1}{0.2} & \valstd{61.3}{1.7} & \valstd{82.8}{0.1} & 73.9 & \gsep & \valstd{62.0}{0.7} & \valstd{67.9}{0.6} & \valstd{56.7}{0.5} & \valstd{67.7}{0.6} & 63.6 \\
 & ORIENT & \valstd{77.4}{0.7} & \valstd{77.6}{0.8} & \valstd{64.4}{0.3} & \valstd{82.6}{0.5} & \underline{75.5} & \gsep & \valstd{64.3}{1.2} & \valstd{69.6}{0.7} & \valstd{59.8}{0.3} & \valstd{68.4}{0.3} & 65.5 \\
\midrule
\multirow{4}{*}{LADA~\citep{sun2023lada}} & Random & \valstd{74.6}{0.9} & \valstd{77.5}{0.1} & \valstd{62.3}{3.3} & \valstd{82.6}{0.2} & 74.2 & \gsep & \valstd{67.5}{0.9} & \valstd{68.0}{0.2} & \valstd{60.0}{0.6} & \valstd{69.1}{0.4} & 66.1 \\
 & CoreSet & \valstd{75.6}{0.2} & \valstd{77.1}{1.1} & \uvalstd{65.9}{1.7} & \valstd{83.0}{0.1} & 75.4 & \gsep & \valstd{66.1}{0.2} & \valstd{67.5}{0.4} & \valstd{59.5}{0.2} & \valstd{67.4}{0.3} & 65.1 \\
 & LESS & \valstd{73.3}{0.4} & \valstd{76.7}{0.1} & \valstd{65.7}{1.5} & \valstd{81.5}{0.5} & 74.3 & \gsep & \valstd{60.5}{1.4} & \valstd{66.2}{0.3} & \valstd{59.1}{0.2} & \valstd{67.1}{0.7} & 63.2 \\
 & ORIENT & \valstd{74.1}{0.5} & \valstd{77.8}{0.2} & \valstd{63.9}{0.1} & \valstd{83.3}{0.1} & 74.8 & \gsep & \valstd{67.9}{0.1} & \valstd{67.5}{0.6} & \valstd{56.9}{1.1} & \valstd{66.2}{0.3} & 64.6 \\
\midrule
\multicolumn{2}{l}{FDA~\citep{yang2020fda}} & \valstd{72.1}{1.9} & \valstd{73.5}{0.2} & \valstd{56.7}{0.8} & \valstd{79.4}{0.1} & 70.4 & \gsep & \valstd{62.4}{0.4} & \valstd{70.1}{1.0} & \valstd{55.7}{3.3} & \valstd{67.1}{2.8} & 63.8 \\
\multicolumn{2}{l}{MIC~\citep{hoyer2023mic}} & \valstd{74.1}{2.2} & \valstd{77.5}{0.3} & \valstd{61.3}{0.7} & \valstd{82.7}{0.2} & 73.9 & \gsep & \valstd{63.6}{0.5} & \valstd{70.1}{0.5} & \valstd{58.6}{0.5} & \valstd{68.9}{0.4} & 65.3 \\
\multicolumn{2}{l}{Pixel Contrastive~\citep{alonso2021pixel} (SSDA)} & \valstd{67.6}{2.2} & \valstd{73.7}{0.6} & \valstd{54.1}{2.0} & \valstd{81.0}{0.6} & 69.1 & \gsep & \uvalstd{69.7}{0.2} & \valstd{64.8}{1.2} & \valstd{58.0}{1.8} & \valstd{65.7}{0.6} & 64.6 \\
\multicolumn{2}{l}{MADA~\citep{ning2021mada} (Active DA)} & \valstd{74.1}{0.4} & \valstd{71.6}{0.7} & \valstd{58.0}{0.7} & \valstd{80.8}{1.6} & 71.1 & \gsep & \valstd{66.6}{1.3} & \bvalstd{70.5}{1.2} & \valstd{59.6}{0.8} & \valstd{69.3}{1.2} & \underline{66.5} \\
\midrule
\textbf{TAC} & \textbf{TAC} & \bvalstd{79.1}{0.3} & \bvalstd{78.9}{0.3} & \bvalstd{67.6}{0.8} & \uvalstd{83.6}{0.2} & \textbf{77.3} & \gsep & \bvalstd{70.5}{1.4} & \uvalstd{70.3}{0.4} & \bvalstd{62.9}{0.8} & \bvalstd{70.0}{0.5} & \textbf{68.4} \\
\bottomrule
\end{tabularx}
\end{table*}

\paragraph{Overall performance at the primary budget.}
At $(B_s,B_t)=(150,10)$, TAC achieves 77.3\% BraTS Macro Dice
and 68.4\% MAMA-MIA foreground Dice, exceeding Random+Random
by 3.7 and 4.8 points, respectively
(\cref{tab:tac_main_results}).
Among the 20 acquisition--selection baseline combinations,
CLUE+ORIENT performs best on BraTS at 75.5\%, while
CoreSet+Random performs best on MAMA-MIA at 66.4\%.
TAC exceeds these methods by 1.8 and 2.0 Dice points,
respectively.

\paragraph{Comparison with domain adaptation.}
TAC also outperforms all four budget-matched domain-adaptation
baselines on both benchmarks.
The strongest DA methods are MIC on BraTS, reaching 73.9\%,
and MADA on MAMA-MIA, reaching 66.5\%.
TAC improves upon these results by 3.4 and 1.9 Dice points,
respectively.
Adaptation protocols and seed-level results are provided in
Supplementary~\cref*{app:da_protocol,app:da_seed_results}.

\paragraph{Performance across target centers.}
Across the eight held-out target centers, TAC ranks first on six
and second on the remaining two among all budget-matched methods.
On BraTS GBM, Random+ORIENT achieves 83.7\% Dice compared with
83.6\% for TAC. On MAMA-MIA I-SPY1, MADA reaches 70.5\%
compared with 70.3\% for TAC.

\begin{figure}[t]
    \centering
    \includegraphics[width=\columnwidth]{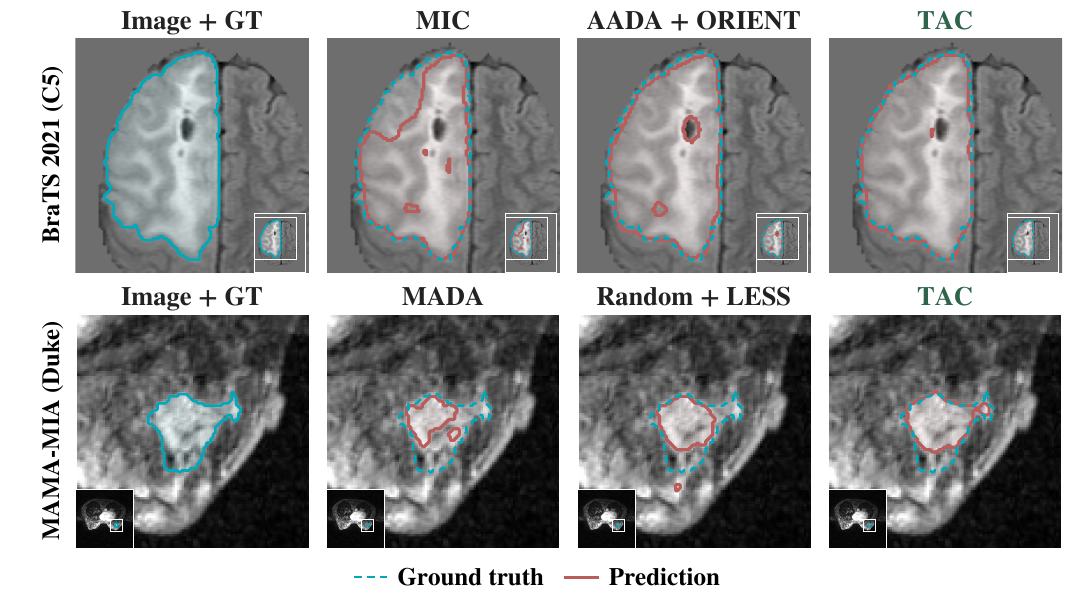}
    \caption{\textbf{Qualitative comparisons}, $(B_s,B_t)=(150,10)$. Top: BraTS C5; bottom: MAMA-MIA Duke. Cyan: ground truth; orange: predictions; insets: full slices.}
    \label{fig:tac_qualitative}
\end{figure}

\paragraph{Qualitative segmentation comparisons.}
\Cref{fig:tac_qualitative} presents representative held-out
cases from BraTS C5 and MAMA-MIA Duke.
Under matched source and target budgets, TAC more closely
follows irregular tumor boundaries and reduces spurious
predicted regions in these examples.

\subsection{Sensitivity to Source and Target Budgets}
\label{subsec:budget_sensitivity}

\begin{figure}[t]
    \centering
    \includegraphics[width=\columnwidth]{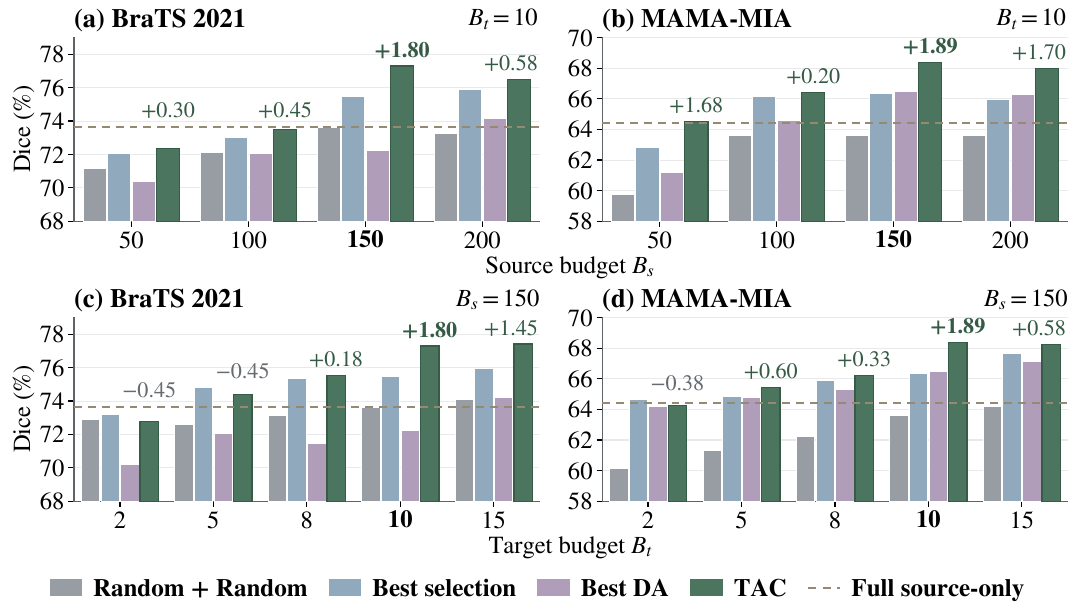}
    \caption{\textbf{Budget sensitivity} on BraTS and MAMA-MIA. Top: vary $B_s$, $B_t=10$; bottom: vary $B_t$, $B_s=150$. Bars: three-seed, four-center means. Best selection: strongest plotted selection-only method; Best DA: strongest budget-matched adaptation method. Annotations: TAC minus the stronger comparator. Dashed: Full Source-only (all source labels, no target labels; non-budget-matched).}
    \label{fig:tac_budget_sweeps}
\end{figure}

\paragraph{Source-budget sensitivity.}
Panels (a,b) of \Cref{fig:tac_budget_sweeps} vary the source
budget used for final training while fixing $B_t=10$.
TAC achieves higher mean Dice than the strongest plotted
budget-matched comparator at all four source budgets on
both benchmarks, although the margins vary.
On BraTS, the gain reaches 1.80 Dice points at $B_s=150$.
On MAMA-MIA, it ranges from 0.20 points at $B_s=100$
to 1.89 at $B_s=150$ and remains 1.70 at $B_s=200$.

\paragraph{Target-budget sensitivity.}
Panels (c,d) fix $B_s=150$ and vary the target annotation
budget over $B_t\in\{2,5,8,10,15\}$.
TAC trails the strongest plotted comparator at $B_t=2$
on both benchmarks and at $B_t=5$ on BraTS.
From $B_t=8$ onward, TAC achieves higher mean Dice on
both benchmarks, with its largest margins of 1.80 and
1.89 points at $B_t=10$ on BraTS and MAMA-MIA, respectively.
At $B_t=15$, the margins remain positive at 1.45 and
0.58 points, respectively.

\subsection{Source--Target Coupling Analysis}
\label{subsec:coupling_main}

\paragraph{Controlled factorial comparison.}
To examine how target acquisition interacts with source curation,
we evaluate a matched $2\times2$ design crossing Random and TAC
acquisition with Random and TAC source selection. We further test
four hybrids that replace either TAC component with CoreSet, CLUE,
or ORIENT. All arms use the same budgets
$(B_s,B_t)=(150,10)$ and final-training protocol. For each arm
using TAC curation, reliability and coverage are recomputed from
the support acquired by its target-selection policy
(\cref*{app:coupling_details}).

\begin{table}[t]
    \centering
    \caption{\textbf{Source--target coupling and hybrid controls.} Three-seed mean $\pm$ sample std of four-center Dice (\%) at $(B_s,B_t)=(150,10)$. $\Delta_{\mathrm{int}}$ is the matched-seed factorial interaction.}
    \label{tab:coupling_main}
    \footnotesize
    \setlength{\tabcolsep}{3.0pt}
    \renewcommand{\arraystretch}{1.13}
    \begin{tabular}{@{}llcc@{}}
        \toprule
        \shortstack[l]{Target\\acquisition}
        & \shortstack[l]{Source\\selection}
        & \shortstack{BraTS\\Avg.}
        & \shortstack{MAMA-MIA\\Avg.} \\
        \midrule
        Random & Random
            & \valstd{73.63}{0.95}
            & \valstd{63.60}{0.55} \\
        TAC & Random
            & \valstd{73.02}{1.14}
            & \valstd{64.27}{0.59} \\
        Random & TAC
            & \valstd{75.11}{0.31}
            & \valstd{64.29}{0.86} \\
        TAC & TAC
            & \valstd{77.31}{0.28}
            & \valstd{68.39}{0.58} \\
        \addlinespace[2pt]
        \multicolumn{2}{l}{Interaction $\Delta_{\mathrm{int}}$}
            & \valstd{+2.82}{0.84}
            & \valstd{+3.44}{1.46} \\
        \midrule
        TAC & CoreSet & \valstd{74.61}{0.03} & \valstd{63.99}{0.33} \\
        CoreSet & TAC & \valstd{74.69}{0.33} & \valstd{64.32}{0.75} \\
        TAC & ORIENT & \valstd{75.40}{0.52} & \valstd{64.63}{0.26} \\
        CLUE & TAC & \valstd{74.49}{0.30} & \valstd{65.24}{0.24} \\
        \bottomrule
    \end{tabular}
\end{table}

\paragraph{Conditional gains and interaction.}
With Random source selection, TAC acquisition decreases Dice by
0.6 points on BraTS but improves it by 0.7 on MAMA-MIA. With
TAC source curation, TAC acquisition instead improves Dice by
2.2 and 4.1 points, respectively.

For each seed, let $D_{ij}$ denote the four-center mean Dice,
where $i$ indexes acquisition and $j$ indexes source selection;
$0$ and $1$ indicate Random and TAC, respectively. We quantify
their interaction as
\begin{equation}
\Delta_{\mathrm{int}}
= D_{11}-D_{10}-D_{01}+D_{00}.
\label{eq:coupling_interaction}
\end{equation}
All three matched-seed interactions are positive on both datasets,
averaging 2.82 and 3.44 Dice points on BraTS and
MAMA-MIA, respectively
(\cref*{tab:coupling_main,tab:coupling_brats_seeds,tab:coupling_mamamia_seeds}).

\paragraph{Strong hybrid controls.}
Full TAC exceeds all four hybrids in every seed on both datasets.
With TAC acquisition fixed, TAC curation outperforms
ORIENT by 1.91 and 3.76 Dice points on BraTS and MAMA-MIA;
the corresponding gaps to CoreSet are 2.70 and 4.40 points.
With TAC curation fixed, TAC acquisition outperforms CLUE by
2.82 and 3.15 points and CoreSet by 2.62 and 4.07 points,
respectively. These comparisons support the benefit of coupling
target acquisition with target-anchored source curation.

\subsection{Component Ablation Studies}
\label{subsec:component_ablation}

\begin{table}[t]
    \centering
    \caption{\textbf{Component ablations} at $(B_s,B_t)=(150,10)$. Three-seed, four-center mean Dice (\%); $\Delta$ is the difference from full TAC. All variants share the acquired target support and training protocol. Per-center results in \cref{tab:ablation_brats_centers,tab:ablation_mamamia_centers}.}
    \label{tab:component_ablation}
    \footnotesize
    \setlength{\tabcolsep}{2.3pt}
    \renewcommand{\arraystretch}{1.12}
    \begin{tabular}{@{}lcccc@{}}
        \toprule
        & \multicolumn{2}{c}{BraTS} & \multicolumn{2}{c}{MAMA-MIA} \\
        \cmidrule(lr){2-3}\cmidrule(l){4-5}
        Variant & Avg. & $\Delta$ & Avg. & $\Delta$ \\
        \midrule
        \textbf{Full TAC} & \textbf{77.3} & 0.0 & \textbf{68.4} & 0.0 \\
        \midrule
        Image-only curation ($M$ only) & 74.5 & -2.8 & 64.3 & -4.1 \\
        \addlinespace[2pt]
        Reliability only & 74.5 & -2.8 & 65.0 & -3.4 \\
        Coverage only (PRISM) & 75.3 & -2.0 & 64.2 & -4.2 \\
        w/o target cap & 74.2 & -3.1 & 65.8 & -2.6 \\
        \addlinespace[2pt]
        Equal $M/G/P$ weights & 74.2 & -3.1 & 65.1 & -3.3 \\
        w/o distribution match ($M$) & 74.9 & -2.4 & 65.3 & -3.1 \\
        w/o gradient agreement ($G$) & 74.7 & -2.6 & 66.6 & -1.7 \\
        w/o prototype ($P$) & 74.4 & -2.9 & 65.3 & -3.1 \\
        \bottomrule
    \end{tabular}
\end{table}

\paragraph{Annotation-informed curation beyond supervision.}
To test whether acquired target annotations improve source
curation beyond their direct supervisory role, we compare
full TAC with an image-only curation control.
Both variants retain the same acquired target support,
support-capped coverage, and final-training protocol.
The control sets $R_j=M_j$, preserving image-derived
distribution matching while removing annotation-derived
gradient agreement and prototype compatibility from source
scoring.
Full TAC improves over this control by 2.8 Dice points on
BraTS and 4.1 points on MAMA-MIA.
Because both variants use the same target masks for final
training, these results demonstrate the additional value
of target annotations for source curation.

\paragraph{Reliability--coverage complementarity.}
We examine the complementary roles of reliability and
coverage by comparing full TAC with two restricted selectors.
Reliability-only selection retains the target-conditioned
scores $R_j$ but selects the top-$B_s$ ranked cases without
coverage optimization.
Coverage-only selection sets $R_j=1$, yielding a PRISM-FLVMI
instantiation~\citep{kothawade2022prism} with relevance scaling $\eta=1$.
We retain the same TAC-acquired target support,
similarity matrices, source budget, and final-training protocol.
Full TAC improves over reliability-only selection by
2.8 and 3.4 Dice points on BraTS and MAMA-MIA, respectively,
and over coverage-only selection by 2.0 and 4.2 points.
These results support combining target-conditioned
reliability with complementary source coverage.

\paragraph{Target-anchored coverage.}
We further examine whether target anchoring is necessary
within the coverage objective.
The target cap limits source coverage according to each
source case's affinity to the acquired target support.
Removing this cap replaces $h_{ij}$ with the source-only
similarity $K_{ij}$ while retaining target-conditioned
reliability and the same final-training protocol.
This reduces average Dice by 3.1 points on BraTS and
2.6 points on MAMA-MIA.
The results indicate that target support contributes not
only to source reliability estimation but also to the
coverage criterion used for subset selection.

\paragraph{Multi-signal reliability and adaptive fusion.}
We compare split-adaptive fusion with equal weighting and remove
each reliability signal in turn.
TAC combines distribution match ($M$), gradient agreement
($G$), and prototype compatibility ($P$) using weights
derived from signal stability and prototype-score spread.
Replacing these weights with uniform fusion reduces
average Dice by 3.1 points on BraTS and 3.3 points on
MAMA-MIA.

\begin{figure}[t]
    \centering
    \includegraphics[width=\columnwidth]{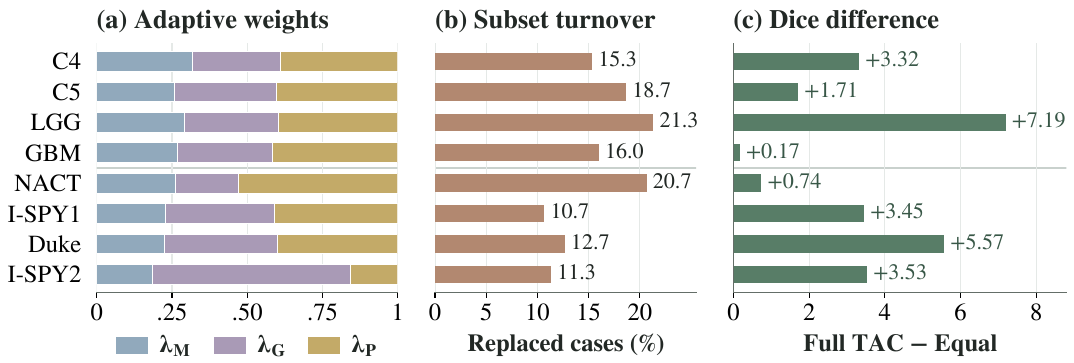}
    \caption{\textbf{Split-adaptive reliability fusion}, $(B_s,B_t)=(150,10)$. (a) Center-specific $M/G/P$ weights. (b) Selected source cases differing from equal weighting (\%). (c) Per-center adaptive minus equal-weight Dice (three-seed means).}
    \label{fig:reliability_fusion}
\end{figure}

\Cref{fig:reliability_fusion} shows that the adaptive weights vary across
held-out centers and change 10.7--21.3\% of selected source cases relative
to equal weighting. Mean Dice improves at all eight centers, by
0.17--7.19 points.

Removing any individual signal reduces average Dice,
with the remaining weights renormalized to sum to one.
The decreases range from 2.4 to 2.9 points on BraTS
and from 1.7 to 3.1 points on MAMA-MIA.
The largest drop occurs when prototype compatibility is
removed on BraTS, while removing either distribution
match or prototype compatibility produces the largest
drop on MAMA-MIA.

\subsection{Computational Cost and Budget Accounting}
\label{subsec:computation}

\begin{table}[t]
    \centering
    \caption{\textbf{TAC computational cost}, $(B_s,B_t)=(150,10)$, one NVIDIA A40. Reconstructed seed-0 GPU-hours, averaged over four target centers; stage costs approximate.}
    \label{tab:compute_main}
    \footnotesize
    \setlength{\tabcolsep}{3pt}
    \renewcommand{\arraystretch}{1.12}
    \begin{tabular}{@{}lrrrr@{}}
        \toprule
        Dataset & Warm-up & Acq.+Evidence & Final & Total \\
        \midrule
        BraTS & 8.809 & 1.872 & 4.957 & \textbf{15.637} \\
        MAMA-MIA & 2.463 & 0.900 & 1.483 & \textbf{4.846} \\
        \bottomrule
    \end{tabular}
\end{table}

Warm-up dominates TAC's end-to-end cost (\cref{tab:compute_main}),
whereas CPU-based source selection takes only 0.73 and 0.80 seconds
per center on BraTS and MAMA-MIA, respectively. MAMA-MIA's
25-epoch warm-up cost is extrapolated from five recorded epochs.
Estimates include training-time validation but exclude annotation
and scheduling overhead. Per-center measurements and accounting
details appear in Supplementary~\cref*{app:compute,tab:compute_tac_e2e}.

\section{Conclusion}
\label{sec:conclusion}

\looseness=-1
We introduced TAC, a target-first framework coupling target acquisition
with source curation for cross-center tumor segmentation. The acquired
support supervises final training and serves as a task-specific reference
for selecting complementary labeled source cases through
reliability-weighted coverage.

With 10 acquired target cases and 150 selected source cases, TAC achieves
77.3\% Macro Dice on BraTS and 68.4\% foreground Dice on MAMA-MIA,
exceeding the strongest budget-matched baselines by 1.8 and 1.9 points,
respectively.
The factorial study further shows a positive interaction between
acquisition and source curation. These findings highlight training-set
composition as a target-specific design decision, informed by the same
annotations that supervise the final segmenter.

\clearpage
\section*{LLM Usage Statement}

OpenAI ChatGPT and Anthropic Claude were used to provide
reviewer-style feedback on manuscript drafts and to assist
with language editing, manuscript organization, related-work
presentation, citation formatting, and refinement of
author-written code. The research ideas, experimental design,
execution, result analysis, and figures were developed and
prepared by the authors without LLM assistance. The authors
reviewed the final content and take full responsibility for
all reported results, citations, and conclusions.

\clearpage
{
    \small
    \setlength{\bibsep}{0pt}
    \bibliographystyle{ieeenat_fullname}
    \bibliography{main}
}

\clearpage
\begingroup

\setcounter{table}{0}
\renewcommand{\thetable}{S\arabic{table}}
\renewcommand{\theHtable}{supp.\arabic{table}}
\setcounter{figure}{0}
\renewcommand{\thefigure}{S\arabic{figure}}
\renewcommand{\theHfigure}{supp.\arabic{figure}}
\setcounter{equation}{0}
\renewcommand{\theequation}{S\arabic{equation}}
\renewcommand{\theHequation}{supp.\arabic{equation}}
\renewcommand{\paragraph}[1]{\par\addvspace{.5em}\noindent\textbf{#1}}

\renewcommand{\valstd}[2]{#1\,{\fontsize{6pt}{7pt}\selectfont$\pm$#2}}
\renewcommand{\bvalstd}[2]{\textbf{#1}\,{\fontsize{6pt}{7pt}\selectfont$\pm$#2}}
\renewcommand{\uvalstd}[2]{\underline{#1}\,{\fontsize{6pt}{7pt}\selectfont$\pm$#2}}

\makeatletter
\setlength{\@fptop}{0pt}
\setlength{\@fpsep}{8pt}
\setlength{\@fpbot}{0pt plus 1fil}
\setlength{\@dblfptop}{0pt}
\setlength{\@dblfpsep}{8pt}
\setlength{\@dblfpbot}{0pt plus 1fil}
\makeatother
\renewcommand{\dblfloatpagefraction}{0.8}
\flushbottom

\renewcommand{\gsep}{}
\clearpage
\maketitlesupplementary

\appendix

\section{Method Implementation Details}
\label{app:method_details}
\label{app:tac_details}

All constants below are shared across target centers.

\subsection{Signal Normalization}
\label{app:normalization}
Each signal to which \(\operatorname{Norm}\) is applied is min--max
normalized over the cases being scored in the current target split,
\[
\operatorname{Norm}(r_i)=\frac{r_i-\min_j r_j}{\max_j r_j-\min_j r_j},
\]
and set to zero when its range is below \(10^{-12}\). Embeddings and gradients
are \(\ell_2\)-normalized with \(\epsilon=10^{-8}\) added to the norm.
The fused reliability \(R_i\) is a convex combination of normalized
\(M_i,G_i,P_i\) and receives no further min--max normalization.

\subsection{Target Acquisition Details}
\label{app:target_acquisition_details}

\paragraph{Slice statistics.}
For each axial slice \(s\), the frozen warm-up model gives pixelwise class
probabilities. TAC records the mean pixel entropy \(\mathrm{ent}^{(s)}\), the
mean foreground probability \(\mathrm{fg}^{(s)}\) (summed over all non-background
classes), the mean class-probability vector \(\bar p^{(s)}\), the globally
average-pooled backbone feature, and the consistency \(C^{(s)}\): the fraction
of pixels whose predicted class is unchanged when the input is scaled by
\(0.85\), scaled by \(1.15\), or shifted by \(+0.03\), averaged over the three
views. All slices of source cases and target candidates are eligible for the same
model-based ranking.

\paragraph{Slice-to-case aggregation.}
Let \(n_j\) be the number of slices in case \(j\), source or target.
Its slices are ranked by
\(\mathrm{fg}^{(s)}+0.25\,\mathrm{ent}^{(s)}\); the top \(\lceil 0.3\,n_j\rceil\)
slices (at least one) are retained, and every statistic is averaged over them.
The averaged feature, \(\ell_2\)-normalized, is the case embedding \(z_j\).
The acquisition signals are \(U^{\mathrm{fg}}_j=\overline{\mathrm{ent}}_j\cdot
\overline{\mathrm{fg}}_j\) and \(C_j=\overline{C}_j\).

\paragraph{Local inconsistency.}
Let \(\mathcal{N}_{10}(j)\) contain the \(\min(10,N_t-1)\) other candidates
with highest cosine similarity to \(z_j\), with ties following candidate order.
The weights and local-inconsistency scores are
\[
\begin{aligned}
w_{jr}&=\frac{[\kappa(z_j,z_r)]_+}
{\sum_{q\in\mathcal{N}_{10}(j)}[\kappa(z_j,z_q)]_++10^{-8}},\\
LI^{0}_j&=1-\textstyle\sum_{r\in\mathcal{N}_{10}(j)}w_{jr}\,\bar p_r^{\top}\bar p_j,\\
LI_j&=LI^{0}_j+\textstyle\sum_{r\in\mathcal{N}_{10}(j)}w_{jr}\,LI^{0}_r .
\end{aligned}
\]
For two mean profiles, \(1-\bar p_r^\top\bar p_j\) is the expected mismatch of independently drawn class labels; \(LI_j\) adds its neighbors' weighted \(LI^0\) scores to its own.
A singleton pool has \(LI_j=0\); otherwise, if all clipped neighbor
similarities are zero, \(LI_j=1\).

\paragraph{Support selection.}
The acquisition weights use \(\omega_{\mathrm{LI}}=0.50\) and
\(\omega_{\mathrm{C}}=0.15\). Weighted \(k\)-means runs on the target embeddings
with sample weights \(\max(a_j,10^{-6})\), \(k=B_t\), \(k\)-means++
initialization, \(10\) restarts, at most \(300\) iterations, and seed \(2025\).
Each cluster \(m\) contributes its member with the largest
\[
\operatorname{Priority}(j;m)=0.7\big(1-\widetilde d_{jm}\big)+0.3\,\widetilde a_j,
\]
where \(\widetilde d_{jm}=\operatorname{Norm}_j(\|z_j-\xi_m\|_2)\) is the
normalized distance to centroid \(\xi_m\) and
\(\widetilde a_j=\operatorname{Norm}(\max(a_j,10^{-6}))\).
Both normalizations range over all candidates, with distances normalized
separately for each centroid; selection is restricted to that cluster's members.
If a cluster has no unused member, the slot is filled by the unused candidate
with the largest \(a_j\). Ties follow the candidate order.

\subsection{Source Reliability}
\label{app:reliability_defaults}
\label{app:checkpoint_details}

\paragraph{Distribution match.}
\(M_i\) uses the \(\min(8,B_t)\) support cases with the largest
\([\kappa(z_i^s,z_r^t)]_+\), breaking ties by fixed support-case order,
and the neighborhood weights
\[
\bar a_{ir}=\frac{\max(a_r,10^{-6})}
{\sum_{q\in\mathcal{N}(i)}\max(a_q,10^{-6})+10^{-12}} .
\]

\paragraph{Gradient checkpoint and extraction.}
Gradient agreement uses the final warm-up checkpoint after one-quarter of the final-training schedule
($e_w=5$ for BraTS; $e_w=25$ for MAMA-MIA). For each case,
the training loss
\(\mathcal{L}=0.5\,\mathcal{L}_{\mathrm{CE}}+0.5\,\mathcal{L}_{\mathrm{Dice}}\)
(Dice smoothing \(10^{-5}\)) is summed over the case's slices in batches of
four and back-propagated to the final segmentation-head parameters only, with
the network in training mode. The accumulated gradient is divided by the number
of slices and \(\ell_2\)-normalized, giving \(g_i^{(e_w)}\) for source case
\(i\). The support gradient \(\bar g^{(e_w)}\) is the
renormalized mean of the normalized support-case gradients.

\paragraph{Prototype compatibility.}
During BraTS preprocessing, we remap the raw enhancing-tumor label from \(4\) to \(3\).
The resulting label masks give the whole-tumor fraction
\(f_{\mathrm{WT}}\) (label \(>0\)), the tumor-core fraction
\(f_{\mathrm{TC}}\) (labels \(1\) and \(3\)), the enhancing-tumor fraction
\(f_{\mathrm{ET}}\) (label \(3\)), and \(h=\mathbf{1}[f_{\mathrm{ET}}>0]\).
For each feature \(f\), support extrema \(f_{\min},f_{\max}\) define
scaling for both source and support cases:
\[
\tilde f=\operatorname{clip}_{[0,1]}\!\left(
\frac{f-f_{\min}}{f_{\max}-f_{\min}+10^{-8}}\right),
\]
giving
\(m=\tilde f_{\mathrm{WT}}+\tilde f_{\mathrm{TC}}+1.5\,\tilde f_{\mathrm{ET}}
+0.5\,\tilde h\).

For MAMA-MIA, the score is the tumor fraction
\[
m(y)=\frac{|\{v\in\Omega_y:y(v)>0\}|}{|\Omega_y|},
\]
where \(\Omega_y\) contains all voxels of the complete preprocessed case
volume, including background, before slice selection or training-patch sampling.
Source and target cases follow the same preprocessing protocol. The score
measures tumor extent relative to this field of view.

For either score, let \(q_1\le q_2\) be the linearly interpolated
\(1/3\) and \(2/3\) quantiles of the support-case scores \(m\). The assignment
\(c(m)=\mathbf{1}[m\ge q_1]+\mathbf{1}[m\ge q_2]\) gives up to three
classes. Source cases use these support-derived thresholds. Each nonempty
support class defines a prototype \(\pi_c\), the normalized mean of its
support embeddings. A source case in class \(c(i)\) receives
\(P^{0}_i=[\kappa(z_i^s,\pi_{c(i)})]_+-\max_{c\neq c(i)}[\kappa(z_i^s,\pi_c)]_+\),
where the maximum ranges over other nonempty support classes.
When its class has no prototype,
\(P^{0}_i=[\kappa(z_i^s,\mu_{\mathcal{T}})]_+\), where \(\mu_{\mathcal{T}}\)
is the normalized support mean. The competing-prototype term is zero when
the support represents a single class. Then \(P_i=\operatorname{Norm}(P^{0}_i)\).

\paragraph{Reliability weights.}
For a score \(r\) and recomputed variants \(\{r^{(v)}\}\), the stability is
\[
\operatorname{stab}(r)=\frac{1}{2V}\sum_{v=1}^{V}\Big[
\tfrac12\big(\rho(r,r^{(v)})+1\big)+o_k(r,r^{(v)})\Big].
\]
Here \(\rho\) is Pearson correlation, clipped to \([-1,1]\) and set to zero
when the product of centered vector norms is below \(10^{-12}\). The top-\(k\) overlap is
\(o_k(r,r')=|\mathrm{Top}_{k}(r)\cap\mathrm{Top}_{k}(r')|/k\), with
\(k=\min(150,N_s)\) and ties following source-case order. The weighting scores are based on support-resampling stability for both
\(M\) and \(G\). We draw eight subsets of
\(\max(2,\operatorname{round}(0.8B_t))\) support cases uniformly without
replacement (seed \(2025\)), recomputing both source-case scores for each
subset; \(G\) uses the corresponding subset-averaged support gradient.
Define
\[
\begin{aligned}
s_M&=\begin{cases}
1, & B_t\le2,\\
\operatorname{stab}(M), & B_t>2,
\end{cases}\\
s_G&=\operatorname{stab}(G),\\
s_P&=0.5+0.5\,\operatorname{std}(P),\\
\boldsymbol{\lambda}&=\operatorname{softmax}(\mathbf{s}/\tau),
\quad \tau=0.3,
\end{aligned}
\]
where \(\mathbf{s}=(s_M,s_G,s_P)\),
\(\boldsymbol{\lambda}=(\lambda_M,\lambda_G,\lambda_P)\), and
\(\operatorname{std}(P)\) is the population standard deviation across
source cases. Resampling stability favors evidence robust to individual
target cases, while prototype-score spread serves as a proxy for source-case
discrimination. The fixed temperature sharpens relative reliability
differences while preserving convex fusion.

\subsection{\texorpdfstring{Reliability-Weighted\\Facility-Location Selection}{Reliability-Weighted Facility-Location Selection}}
\label{app:unified_selector}

\paragraph{Target-capped similarities.}
For each source case $i$, define
\[
\begin{aligned}
K_{ij}&=[\kappa(z_i^s,z_j^s)]_+,\\
c_i&=\max_{q\in\mathcal{T}}[\kappa(z_i^s,z_q^t)]_+,\\
h_{ij}&=\min(K_{ij},c_i).
\end{aligned}
\]
The cap $c_i$ applies to the represented source case $i$ and limits the contribution of
source regions with weak support affinity; only $c_i=0$ removes
that region's contribution entirely. The unweighted coverage is
$C(\mathcal{S};\mathcal{T})=\sum_i\max_{j\in\mathcal{S}}h_{ij}$,
with empty-set maxima defined as zero. Since $c_i$ is independent of the
selected representative $j$,
\[
\begin{aligned}
C(\mathcal{S};\mathcal{T})
&=\sum_i\min\!\left(\max_{j\in\mathcal{S}}K_{ij},\,c_i\right).
\end{aligned}
\]
This is the FLVMI form in PRISM~\citep{kothawade2022prism}, with source
cases as the ground set, acquired target cases as the query set, and
relevance scaling fixed to one.

\paragraph{Reliability-weighted objective.}
The direct convex fusion
$R_j=\lambda_M M_j+\lambda_G G_j+\lambda_P P_j$
lies in $[0,1]$ and is used directly.
The utility of representing case $i$ by case $j$ is $v_{ij}=R_jh_{ij}$.
The resulting objective remains standard facility location on a modified
utility matrix; TAC specifies this matrix through its annotation-conditioned
reliability construction and acquired target reference.
The source-selection objective is
\[
F(\mathcal{S};\mathcal{T})=
\sum_{i\in\mathcal{D}_s}\max_{j\in\mathcal{S}}R_jh_{ij}.
\]
The source budget imposes $|\mathcal{S}|=B_s$.
Reliability weights each candidate representative inside the maximum.
Each source case contributes only its best reliability-weighted similarity;
additional representatives are rewarded only where they improve this value.

\paragraph{Greedy selection and caching.}
For $0\le B_s\le N_s$, initialize $\mathcal{S}=\varnothing$ and
$m_i=0$ for all source cases. At each step, compute
\[
\Delta_jF(\mathcal{S})=
\sum_{i\in\mathcal{D}_s}[R_jh_{ij}-m_i]_+,
\]
for each unselected candidate $j$.
Add a candidate $j^\star$ with maximum gain and update
\[
m_i\leftarrow\max(m_i,R_{j^\star}h_{ij^\star})
\quad\text{for all }i.
\]
The cache then equals
$m_i(\mathcal{S})=\max_{j\in\mathcal{S}}R_jh_{ij}$.
Ties are resolved by larger $R_j$, then fixed source-case order.
Selection continues to $|\mathcal{S}|=B_s$, including steps where
all marginal gains are zero.

\paragraph{Degenerate and uniform scores.}
If all $R_j=0$, then $F\equiv0$ and the fixed source-case order
determines selection.
If all $h_{ij}=0$, gains are zero and the tie rule selects by
reliability, then source-case order. With uniform reliability
$R_j=r>0$, $F=rC$ and greedy selection reduces to coverage selection
with the same fixed-order tie rule. In particular, $R_j=1$ for
every source case recovers $C$ exactly.

\paragraph{Diminishing returns and guarantee.}
Write $v_{ij}=R_jh_{ij}\ge0$, fixed before selection.
For $\mathcal{A}\subseteq\mathcal{B}$,
$m_i(\mathcal{A})\le m_i(\mathcal{B})$, and hence, for
$j\notin\mathcal{B}$,
\[
[v_{ij}-m_i(\mathcal{A})]_+
\ge [v_{ij}-m_i(\mathcal{B})]_+.
\]
Summing over $i$ proves diminishing returns. All marginal gains
are nonnegative, so $F$ is monotone and submodular with
$F(\varnothing)=0$. Exact greedy selection under the cardinality
budget therefore satisfies~\citep{nemhauser1978analysis}
\[
F(\mathcal{S}_{\mathrm{greedy}};\mathcal{T})
\ge (1-1/e)F(\mathcal{S}^{\star};\mathcal{T}),
\]
where $\mathcal{S}^{\star}$ maximizes $F$ over $|\mathcal{S}|\le B_s$.
The guarantee applies to $F$. Positive uniform rescaling of all $R_j$ rescales $F$
and its gains without changing the greedy choices.

\section{Experimental Protocol and Reproducibility}
\label{app:experiment_details}

\subsection{Patient-Disjoint Cross-Center Splits}
\label{app:split_details}
Throughout this study, a center denotes a held-out evaluation domain. For BraTS, domains are defined by collection, and for MAMA-MIA, by cohort. A domain may include more than one institution.
Each center retains its predefined target test set. Approximately 15\% of the total target cases (minimum five, without an upper cap) are allocated to validation from the non-test cases; the remaining non-test cases form the initially unlabeled acquisition pool. Splits are patient-disjoint and fixed across methods, budgets, and training seeds. The budget $B_t$ counts newly annotated target training cases from this pool; pre-held-out validation labels are outside $B_t$. Validation serves only for checkpoint selection and is excluded from acquisition, reference construction, reference-gradient computation, and training; the test set is used only for final evaluation. Table~\ref{tab:tac_protocol_splits} reports the exact split sizes.

\begin{table*}[t]
\centering
\caption{Patient-disjoint target splits. Cand.: unlabeled acquisition pool; Val./Test: excluded from acquisition and training; Full src.: labeled cases from other centers. Budget columns: annotated fraction $100B_t/\mathrm{Cand.}$ (\%).}
\label{tab:tac_protocol_splits}
\label{tab:data_splits}
\scriptsize
\setlength{\tabcolsep}{4pt}
\renewcommand{\arraystretch}{1.08}
\begin{tabular}{@{}llrrrrrrrrr@{}}
\toprule
Dataset & Target & Cand. & Val. & Test & Full src. & \multicolumn{5}{c}{Target annotations as fraction of Cand. (\%)} \\
\cmidrule(lr){7-11}
& & & & & & $B_t=2$ & $B_t=5$ & $B_t=8$ & $B_t=10$ & $B_t=15$ \\
\midrule
\multirow{4}{*}{BraTS} & C4 & 15 & 5 & 10 & 1221 & 13.3\% & 33.3\% & 53.3\% & 66.7\% & 100.0\% \\
 & C5 & 26 & 7 & 14 & 1204 & 7.7\% & 19.2\% & 30.8\% & 38.5\% & 57.7\% \\
 & LGG & 35 & 10 & 20 & 1186 & 5.7\% & 14.3\% & 22.9\% & 28.6\% & 42.9\% \\
 & GBM & 55 & 15 & 32 & 1149 & 3.6\% & 9.1\% & 14.5\% & 18.2\% & 27.3\% \\
\midrule
\multirow{4}{*}{MAMA-MIA} & NACT & 19 & 10 & 35 & 1442 & 10.5\% & 26.3\% & 42.1\% & 52.6\% & 78.9\% \\
 & I-SPY1 & 40 & 26 & 105 & 1335 & 5.0\% & 12.5\% & 20.0\% & 25.0\% & 37.5\% \\
 & Duke & 63 & 44 & 184 & 1215 & 3.2\% & 7.9\% & 12.7\% & 15.9\% & 23.8\% \\
 & I-SPY2 & 172 & 147 & 661 & 526 & 1.2\% & 2.9\% & 4.7\% & 5.8\% & 8.7\% \\
\bottomrule
\end{tabular}
\end{table*}

\subsection{Checkpoint Selection and Seed Reporting}
\label{app:reporting_protocol}
All budgeted methods, including FDA, MIC, pixel-level contrastive SSDA, and MADA, are evaluated over three independent training seeds. For each seed $r\in\{0,1,2\}$, TAC and acquisition--selection baselines independently train the selection-time warm-up, recompute target acquisition and source selection, and train the final segmenter. Fixed internal random states (e.g., 2025 for clustering) make these steps reproducible without replacing the training seed. After each epoch, we evaluate $\mathcal{L}_{\mathrm{val}}=0.5\mathcal{L}_{\mathrm{Dice}}+0.5\mathcal{L}_{\mathrm{CE}}$ on the fixed target validation split. Across methods and seeds, we select the checkpoint with the lowest validation loss. Test Dice is used only for final evaluation. For each seed, the dataset-level result is the unweighted mean of the four held-out centers. For the main table, per-center means and sample standard deviations are computed across the three seeds; the four-center Avg.\ reports only the mean of the three seed-level dataset averages. Source- and target-budget curves display only the three-seed means.

\subsection{Acquisition--Selection Baselines}
\label{app:baseline_protocol}
\paragraph{Case-level protocol.}
Both budgets count whole patients rather than 2D slices. At each $(B_s,B_t)$, an acquisition policy selects $B_t$ cases from the initially unlabeled \texttt{target\_train} pool; their segmentation masks are accessed only after selection. Slice embeddings and prediction statistics are aggregated to the patient level before ranking or clustering. A source selector then chooses $B_s$ already-labeled cases from the other centers. All acquisition--selection pairs use matched splits, budgets, and final supervised training.

\paragraph{Target acquisition.}
We evaluate \textbf{Random} (uniform sampling without replacement), \textbf{CoreSet}~\citep{sener2018coreset} (greedy $k$-center on $\ell_2$-normalized patient embeddings), \textbf{AADA}~\citep{su2020aada} (patient-aggregated prediction entropy weighted by $(1-p_s)/p_s$, using source-domain probability $p_s$ from a DANN discriminator), \textbf{CLUE}~\citep{prabhu2021clue} (entropy-weighted embedding $k$-means with $k=B_t$ and distinct centroid-nearest queries), and \textbf{LADA}~\citep{sun2023lada} (prediction disagreement smoothed over $K=10$ neighbors, followed by clustering the top $\min(11B_t,N_t)$ candidates to query $B_t$ cases). These acquisition-specific models supply selection scores; the final segmenter is trained separately.

\paragraph{Source selection.}
Each acquisition is paired with \textbf{Random} (uniform sampling without replacement), \textbf{CoreSet} (source-only greedy $k$-center coverage), \textbf{LESS}~\citep{xia2024less}, and \textbf{ORIENT}~\citep{karanam2022orient}. Random and CoreSet do not depend on the acquired target support, so their source lists can be shared across acquisition arms for a fixed split, budget, and seed. Our Adam-aware head-gradient adaptation of LESS uses source-only warm-up Dice--cross-entropy head gradients, Adam-moment-adjusted source directions, and ordinary gradients on the acquired target support. Both are projected to 256 dimensions and normalized; support-averaged cosine scores are aggregated across available checkpoints with learning-rate weights. Our static ORIENT adaptation scores frozen source embeddings against the acquired target embeddings once before training: BraTS uses nonnegative cosine facility-location mutual information with lazy greedy, while MAMA-MIA uses target-coverage gains with a $0.08$ source-redundancy penalty. LESS and ORIENT are recomputed for each acquisition support set; neither reselects cases during final training.

\paragraph{Final supervision and references.}
The selected patient lists enter standard supervised training: ADE20K-initialized four-channel EfficientViT-Seg-L1 for 20 epochs on BraTS and three-channel 2D nnU-Net v2~\citep{isensee2021nnunet,isensee2024nnunetrevisited} from scratch for 100 epochs on MAMA-MIA (\cref{app:brats_training,app:mamamia_training}). Selection-time evidence and final segmentation training are separate stages (\cref{app:selection_time_protocol}). The reference block distinguishes Full Source-only ($B_t=0$, all labeled source cases), Random + Full Source ($B_t=10$, all labeled source cases), and Random target-only ($B_t=10$, no source cases). These references are outside the matched $B_s=150$ comparisons and are excluded from budget-matched rankings.

\subsection{BraTS Training Protocol}
\label{app:brats_training}
\paragraph{Preprocessing and model.}
BraTS uses nnU-Net preprocessing followed by 2D axial-slice training. Four MRI modalities are retained per slice, independently min--max normalized to $[0,1]$, and resized to $512\times512$. The final segmentation backbone is EfficientViT-Seg-L1 initialized from ADE20K-pretrained weights, with four input channels and four output classes.

\paragraph{Training and evaluation.}
We train for 20 epochs with batch size 4, AdamW (initial learning rate $10^{-4}$, weight decay $10^{-5}$), cosine decay to $10^{-6}$, and a combined Dice--cross-entropy objective. Case-level predictions are reconstructed into volumes for 3D ET/TC/WT evaluation.

\subsection{MAMA-MIA Training Protocol}
\label{app:mamamia_training}
\paragraph{Model and training.}
MAMA-MIA uses nnU-Net v2's 2D PlainConvUNet, initialized from scratch, with three DCE input channels, two output classes (background and tumor), and seven feature stages with $32/64/128/256/512/512/512$ channels. We use SGD with momentum $0.99$, Nesterov acceleration, initial learning rate $0.01$, weight decay $3\times10^{-5}$, and polynomial learning-rate decay with exponent $0.9$. The supervised objective equally weights Dice and cross-entropy with deep supervision; foreground batch Dice uses smoothing $10^{-5}$. The network uses InstanceNorm2d and foreground oversampling probability $0.33$ for labeled training patches.

\paragraph{Preprocessing and augmentation.}
DCE inputs are normalized independently by channel-wise Z-scores (\texttt{use\_mask\_for\_norm=false}); the in-plane resampling spacing is approximately $0.7031\times0.7031$~mm. We use the standard nnU-Net spatial and appearance augmentations: rotation, scaling, mirroring, noise, blur, brightness, contrast, gamma, and simulated low resolution. The three input channels are identified as
\texttt{dce\_0000/0001/0002}.

\paragraph{Inference.}
At inference, a 2D sliding window covers the full case with 50\% overlap and Gaussian weighting, without mirroring/test-time augmentation.

\subsection{Selection-Time Warm-Up}
\label{app:selection_time_protocol}
TAC uses a full-source warm-up lasting one-quarter of the final-training
schedule: five epochs for BraTS and 25 for MAMA-MIA.
Selection evidence is extracted from the final warm-up checkpoint;
final segmentation training uses an independently initialized model.
The source budget $B_s$ limits the selected cases in final training.

\subsection{Computational Cost}
\label{app:compute}

\paragraph{Measured full-source and final-training stages.}
All full-source and final segmentation runs in
\cref{tab:compute_training} used one NVIDIA A40 GPU. Each value is a
measured GPU-hour cost for the specified held-out center; the
Avg.\ row is the unweighted four-center mean. The full-source column measures the complete source-training schedule
(20 epochs for BraTS and 100 for MAMA-MIA). The two final-training columns use
$(B_s,B_t)=(150,10)$. These A40 TAC timings cover final training only; warm-up,
evidence extraction, and subset selection are excluded. The separate
A40 seed-0 end-to-end accounting below includes these stages.

\begin{table}[t]
    \centering
    \caption{\textbf{Full-source and final-training GPU-hours} by target on NVIDIA A40. Final training: $(B_s,B_t)=(150,10)$. Excludes selection and evidence extraction.}
    \label{tab:compute_training}
    \scriptsize
    \setlength{\tabcolsep}{3pt}
    \renewcommand{\arraystretch}{1.1}
    \begin{tabular}{@{}llrrr@{}}
        \toprule
        Dataset & Center & Full-source & R+R final & TAC final \\
        \midrule
        \multirow{5}{*}{BraTS} & C4 & 27.64 & 3.45 & 3.50 \\
        & C5 & 26.75 & 3.60 & 3.42 \\
        & LGG & 24.56 & 3.41 & 3.32 \\
        & GBM & 22.39 & 3.44 & 3.40 \\
        \cmidrule(l){2-5}
        & Avg. & 25.34 & 3.48 & 3.41 \\
        \midrule
        \multirow{5}{*}{MAMA-MIA} & NACT & 16.85 & 1.50 & 1.51 \\
        & I-SPY1 & 15.97 & 1.49 & 1.52 \\
        & Duke & 9.72 & 1.54 & 1.49 \\
        & I-SPY2 & 4.62 & 1.50 & 1.51 \\
        \cmidrule(l){2-5}
        & Avg. & 11.79 & 1.51 & 1.51 \\
        \bottomrule
    \end{tabular}
\end{table}

\paragraph{TAC end-to-end cost reconstructed from cluster stage logs.}
\Cref{tab:compute_tac_e2e} reports a separate, single-A40, seed-0
accounting at $(B_s,B_t)=(150,10)$. Unlike the final-training-only \cref{tab:compute_training},
it includes the full-source warm-up, target acquisition and evidence
extraction, subset-selection latency, and final segmentation training
with validation. BraTS uses five warm-up epochs and 20 final-training
epochs; MAMA-MIA uses 25 warm-up epochs and 100 final-training
epochs. The MAMA-MIA warm-up column therefore charges five times
the duration of a five-epoch block. Selection itself runs on CPU
and is reported in seconds.
End-to-end GPU-hours reflect the reconstructed active A40 allocation
and include the brief selection latency; queue time, failed retries,
and post-training test-checkpoint evaluation are excluded. Stage
components are rounded and thus need not reproduce the totals at
the last decimal; Avg.\ is the equally weighted mean over four held-out
centers. These timings are a single-seed measurement, not a three-seed
cost estimate, and should not be interpreted as the cost of the separate
final-training-only or domain-adaptation runs.

\begin{table*}[t]
    \centering
    \caption{\textbf{TAC end-to-end runtime}, seed 0, one NVIDIA A40, $(B_s,B_t)=(150,10)$, reconstructed from cluster logs. GPU stages: approximate GPU-hours (final training includes validation); CPU selection: seconds. Warm-up: five epochs for BraTS; MAMA-MIA's 25 epochs are extrapolated as five times its five-epoch duration.}
    \label{tab:compute_tac_e2e}
    \scriptsize
    \setlength{\tabcolsep}{4pt}
    \renewcommand{\arraystretch}{1.10}
    \begin{tabular}{@{}llrrrrr@{}}
        \toprule
        Dataset & Target & Warm-up & Acquisition + evidence & Selection (CPU s) & Final training & End-to-end \\
        \midrule
        \multirow{5}{*}{BraTS} & C4 & 9.045 & 1.914 & 0.91 & 4.991 & 15.950 \\
        & C5 & 8.912 & 1.886 & 0.67 & 4.929 & 15.727 \\
        & LGG & 8.769 & 1.822 & 0.78 & 5.026 & 15.617 \\
        & GBM & 8.509 & 1.865 & 0.56 & 4.882 & 15.256 \\
        \cmidrule(l){2-7}
        & Avg. & 8.809 & 1.872 & 0.73 & 4.957 & 15.637 \\
        \midrule
        \multirow{5}{*}{MAMA-MIA} & NACT & 2.750 & 0.961 & 1.43 & 1.478 & 5.189 \\
        & I-SPY1 & 2.500 & 0.916 & 0.99 & 1.485 & 4.901 \\
        & Duke & 2.175 & 0.924 & 0.65 & 1.491 & 4.590 \\
        & I-SPY2 & 2.425 & 0.799 & 0.12 & 1.479 & 4.703 \\
        \cmidrule(l){2-7}
        & Avg. & 2.463 & 0.900 & 0.80 & 1.483 & 4.846 \\
        \bottomrule
    \end{tabular}
\end{table*}

\paragraph{Domain-adaptation training stages.}
\Cref{tab:compute_da_summary} summarizes measured DA training-stage
GPU-hours for the four adaptation methods, including joint source--target
supervision for FDA and MIC. Target-level timings for both benchmarks
appear in \cref{tab:compute_da_brats,tab:compute_da_mamamia}.

\begin{table}[t]
    \centering
    \caption{\textbf{Measured training-stage GPU-hours} for four DA comparators. Avg.: unweighted four-center mean.}
    \label{tab:compute_da_summary}
    \footnotesize
    \setlength{\tabcolsep}{4pt}
    \begin{tabular}{@{}lrr@{}}
        \toprule
        Method & BraTS Avg. & MAMA-MIA Avg. \\
        \midrule
        FDA & 10.74 & 8.72 \\
        MIC & 14.82 & 11.42 \\
        Pixel Contrastive & 9.02 & 8.68 \\
        MADA & 9.11 & 8.63 \\
        \bottomrule
    \end{tabular}
\end{table}

\begin{table}[t]
    \centering
    \caption{\textbf{BraTS DA training-stage GPU-hours} by target. Avg.: unweighted four-center mean.}
    \label{tab:compute_da_brats}
    \scriptsize
    \setlength{\tabcolsep}{2.4pt}
    \renewcommand{\arraystretch}{1.1}
    \begin{tabular}{@{}lrrrrr@{}}
        \toprule
        Method & C4 & C5 & LGG & GBM & Avg. \\
        \midrule
        FDA & 10.93 & 10.56 & 10.59 & 10.88 & 10.74 \\
        MIC & 15.18 & 14.45 & 14.51 & 15.16 & 14.82 \\
        Pixel Contrastive & 9.29 & 8.78 & 8.93 & 9.07 & 9.02 \\
        MADA & 9.32 & 8.90 & 8.91 & 9.32 & 9.11 \\
        \bottomrule
    \end{tabular}
\end{table}

\begin{table}[t]
    \centering
    \caption{\textbf{MAMA-MIA DA training-stage GPU-hours} by target. Avg.: unweighted four-center mean.}
    \label{tab:compute_da_mamamia}
    \scriptsize
    \setlength{\tabcolsep}{2.4pt}
    \renewcommand{\arraystretch}{1.1}
    \begin{tabular}{@{}lrrrrr@{}}
        \toprule
        Method & NACT & I-SPY1 & Duke & I-SPY2 & Avg. \\
        \midrule
        FDA & 8.66 & 8.70 & 8.87 & 8.66 & 8.72 \\
        MIC & 11.16 & 11.09 & 11.99 & 11.44 & 11.42 \\
        Pixel Contrastive & 8.22 & 7.63 & 9.48 & 9.41 & 8.68 \\
        MADA & 7.77 & 7.02 & 10.37 & 9.38 & 8.63 \\
        \bottomrule
    \end{tabular}
\end{table}

\subsection{Domain Adaptation Comparison Protocol}
\label{app:da_protocol}
At each evaluated $(B_s,B_t)$ budget, all four DA baselines use a
shared randomly selected $B_s$-case labeled source subset. FDA, MIC,
and pixel-level contrastive SSDA use the same $B_t$ randomly selected labeled
target cases as Random+Random; MADA actively selects its own $B_t$
target cases. FDA and MIC are trained with \emph{joint source--target
supervision from the first epoch}, while retaining their original
adaptation objectives. Neither includes a separate target-only
fine-tuning stage. Both run for 20 epochs on BraTS and 100 epochs
on MAMA-MIA, matching the total schedule of the supervised controls.

All four methods use additional unlabeled images only from
\texttt{target\_train}; target validation and test partitions are
excluded from selection and adaptation. The benchmark-specific
segmentation backbone and supervised loss are shared, but
method-specific adaptation losses, training stages, and computation
may differ. Label budgets are matched at every point in both
source- and target-budget sweeps.

\begin{itemize}[topsep=3pt,itemsep=2pt,parsep=0pt]
\item \textbf{FDA}~\citep{yang2020fda}: channel-wise 2D Fourier
adaptation replaces the low-frequency source amplitude with target
amplitude while retaining source phase and segmentation masks.
Adapted-source supervision and target entropy minimization are combined
with unit-weight labeled-target Dice--cross-entropy at every update from
epoch one. Total epochs: 20 (BraTS) or 100 (MAMA-MIA).
Hyperparameters: single $\beta=0.01$, entropy weight $0.005$,
and entropy-penalty exponent $2$.
\item \textbf{MIC + DACS}~\citep{hoyer2023mic,tranheden2021dacs}: an EMA teacher
predicts pseudo-labels on unmasked target images. Student training
combines source supervision, source-ground-truth/target-pseudo-label
ClassMix, masked-target consistency, and unit-weight labeled-target
Dice--cross-entropy from epoch one. The pseudo-label loss is weighted
by the fraction of target pixels above the confidence threshold.
Total epochs: 20 (BraTS) or 100 (MAMA-MIA).
Hyperparameters: teacher EMA $0.999$, pseudo-label threshold
$0.968$, mask ratio $0.7$, block size $64$, and consistency weight $1$.
\item \textbf{Pixel-level contrastive SSDA}~\citep{alonso2021pixel}:
a 256-dimensional projection head on decoder features and a class-wise
memory bank, updated by high-confidence, correctly classified labeled
target pixels from an EMA teacher. Source/target supervision is combined
with confidence-weighted pseudo-labeling, entropy regularization, and
class-wise pixel contrastive learning; class frequencies use the sampled
training batches. BraTS recalibrates BatchNorm statistics on unlabeled
\texttt{target\_train} images before inference, whereas MAMA-MIA retains
nnU-Net's InstanceNorm. Continuous training lasts 20 epochs on BraTS
or 100 on MAMA-MIA. Hyperparameters: 256 memory entries per class,
projection dimension 256, contrastive weight $0.2$, entropy weight
$0.01$, 2000-step ramp, and memory confidence threshold $0.95$.
\item \textbf{MADA}~\citep{ning2021mada}: a source-supervised,
output-space adversarial warm-up precedes active querying. Ten anchors
are built from source patient-level class features; the $B_t$ targets
farthest from their nearest anchors are annotated. Stage 1 uses labeled
source and acquired target supervision; stage 2 adds target pseudo-labels
and multi-anchor alignment, starting from the best stage-1 validation
checkpoint with a restarted optimizer and learning-rate schedule.
BraTS uses a 5-epoch warm-up, 5-epoch stage 1, and 10-epoch stage 2,
while MAMA-MIA uses 20 + 20 + 60 epochs. Hyperparameters: 10 anchors,
feature-alignment weight $0.7$, and anchor EMA $0.999$.
\end{itemize}

\subsection{Domain Adaptation Seed-Level Results}
\label{app:da_seed_results}
The adaptation results in Table~2 of the main paper aggregate three
independent training seeds at each target center. \Cref{tab:da_brats_seeds,tab:da_mamamia_seeds} give the individual BraTS macro Dice and MAMA-MIA foreground Dice
values, respectively. Both are reported as case-level Dice percentages;
the main-table means and per-center sample standard deviations are
computed across seeds in the same units. The
four-center Avg.\ is the mean of the three seed-level dataset averages.

\begin{table}[t]
\centering
\caption{\textbf{BraTS DA results by seed}, $(B_s,B_t)=(150,10)$. 3D macro Dice over ET/TC/WT (\%). Table~2 of the main paper reports the three-seed mean $\pm$ sample std. MADA acquires target support; others share ten randomly selected cases.}
\label{tab:da_brats_seeds}
\scriptsize
\setlength{\tabcolsep}{3.1pt}
\renewcommand{\arraystretch}{1.06}
\begin{tabular}{@{}llrrrr@{}}
\toprule
Method & Seed & C4 & C5 & LGG & GBM \\
\midrule
\multirow{3}{*}{FDA} & 0 & 69.99 & 73.44 & 56.49 & 79.33 \\
& 1 & 73.08 & 73.27 & 57.61 & 79.36 \\
& 2 & 73.30 & 73.76 & 56.02 & 79.53 \\
\midrule
\multirow{3}{*}{MIC} & 0 & 71.55 & 77.60 & 60.73 & 82.73 \\
& 1 & 75.53 & 77.78 & 61.06 & 82.41 \\
& 2 & 75.14 & 77.17 & 62.11 & 82.82 \\
\midrule
\multirow{3}{*}{Pixel Contrastive} & 0 & 65.20 & 73.10 & 52.80 & 80.33 \\
& 1 & 69.60 & 74.20 & 53.06 & 81.62 \\
& 2 & 68.10 & 73.90 & 56.35 & 81.04 \\
\midrule
\multirow{3}{*}{MADA} & 0 & 73.70 & 72.40 & 58.40 & 82.60 \\
& 1 & 74.42 & 71.10 & 57.20 & 79.70 \\
& 2 & 74.08 & 71.28 & 58.30 & 79.95 \\
\bottomrule
\end{tabular}
\end{table}

\begin{table}[t]
\centering
\caption{\textbf{MAMA-MIA DA results by seed}, $(B_s,B_t)=(150,10)$. Foreground Dice (\%). Table~2 of the main paper reports the three-seed mean $\pm$ sample std. MADA acquires target support; others share ten randomly selected cases.}
\label{tab:da_mamamia_seeds}
\scriptsize
\setlength{\tabcolsep}{3.1pt}
\renewcommand{\arraystretch}{1.06}
\begin{tabular}{@{}llrrrr@{}}
\toprule
Method & Seed & NACT & I-SPY1 & Duke & I-SPY2 \\
\midrule
\multirow{3}{*}{FDA} & 0 & 62.70 & 71.30 & 52.20 & 70.40 \\
& 1 & 62.50 & 69.40 & 56.40 & 65.60 \\
& 2 & 61.90 & 69.60 & 58.60 & 65.40 \\
\midrule
\multirow{3}{*}{MIC} & 0 & 63.32 & 69.99 & 59.14 & 68.54 \\
& 1 & 64.19 & 69.73 & 58.49 & 68.91 \\
& 2 & 63.29 & 70.62 & 58.14 & 69.37 \\
\midrule
\multirow{3}{*}{Pixel Contrastive} & 0 & 69.54 & 66.16 & 56.47 & 65.04 \\
& 1 & 70.01 & 64.25 & 60.04 & 66.07 \\
& 2 & 69.66 & 64.06 & 57.49 & 66.07 \\
\midrule
\multirow{3}{*}{MADA} & 0 & 67.60 & 69.80 & 58.70 & 68.30 \\
& 1 & 67.00 & 69.90 & 60.30 & 69.00 \\
& 2 & 65.10 & 71.90 & 59.70 & 70.70 \\
\bottomrule
\end{tabular}
\end{table}

\begin{table*}[p]
\centering
\caption{\textbf{BraTS complete budget sweeps} (Dice \%). Source sweeps fix $B_t=10$ and target sweeps fix $B_s=150$. Entries are mean $\pm$ sample std over three seeds. Avg.: unweighted four-center mean.}
\label{tab:budget_sweep_brats_full}
\fontsize{6.5pt}{7.4pt}\selectfont
\setlength{\tabcolsep}{0.9pt}
\renewcommand{\arraystretch}{0.96}
\begin{tabularx}{\textwidth}{@{}l@{\hspace{3pt}\gsep\hspace{3pt}}l*{5}{Y}@{\hspace{3pt}}c@{\hspace{3pt}}*{5}{Y}@{}}
\toprule
\multirow{2}{*}{\shortstack[l]{Target\\acq.}} & \multirow{2}{*}{\shortstack[l]{Source\\sel.}} & \multicolumn{5}{c}{$B_s=50,\ B_t=10$} & \gsep & \multicolumn{5}{c}{$B_s=100,\ B_t=10$} \\
\cmidrule(lr){3-7}\cmidrule(lr){9-13}
 & & C4 & C5 & LGG & GBM & Avg. & \gsep & C4 & C5 & LGG & GBM & Avg. \\
\midrule
\multirow{3}{*}{Random} & Random & \valstd{71.5}{1.6} & \valstd{77.3}{0.4} & \valstd{54.2}{2.7} & \valstd{81.6}{0.9} & 71.2 & \gsep & \valstd{75.3}{1.3} & \valstd{77.5}{0.9} & \valstd{53.2}{2.0} & \valstd{82.6}{0.3} & 72.2 \\
 & CoreSet & \valstd{72.7}{1.0} & \valstd{77.5}{1.0} & \valstd{53.5}{2.0} & \valstd{81.8}{1.0} & 71.4 & \gsep & \valstd{69.2}{0.5} & \valstd{73.1}{0.4} & \valstd{54.7}{1.5} & \valstd{81.5}{0.8} & 69.6 \\
 & ORIENT & \valstd{72.6}{1.1} & \valstd{77.2}{0.7} & \valstd{53.1}{2.4} & \valstd{81.5}{0.7} & 71.1 & \gsep & \valstd{71.5}{1.2} & \valstd{76.6}{0.4} & \valstd{54.9}{2.8} & \valstd{82.6}{0.4} & 71.4 \\
\addlinespace[1pt]
\multirow{3}{*}{CoreSet} & Random & \valstd{73.4}{0.6} & \valstd{77.3}{1.1} & \valstd{52.4}{3.7} & \valstd{81.0}{0.7} & 71.0 & \gsep & \valstd{75.0}{2.2} & \valstd{78.5}{0.6} & \valstd{52.8}{2.6} & \valstd{81.8}{0.5} & 72.0 \\
 & CoreSet & \valstd{73.2}{1.2} & \valstd{77.8}{0.9} & \valstd{53.3}{1.2} & \valstd{82.6}{0.5} & 71.7 & \gsep & \valstd{70.1}{1.5} & \valstd{74.8}{0.6} & \valstd{53.6}{1.3} & \valstd{81.5}{0.7} & 70.0 \\
 & ORIENT & \valstd{72.4}{0.4} & \valstd{77.9}{0.8} & \valstd{54.0}{1.8} & \valstd{81.8}{0.1} & 71.5 & \gsep & \valstd{73.8}{0.8} & \valstd{78.0}{0.3} & \valstd{57.3}{1.8} & \valstd{82.3}{0.7} & 72.9 \\
\addlinespace[1pt]
\multirow{3}{*}{AADA} & Random & \valstd{73.1}{1.9} & \valstd{77.4}{0.6} & \valstd{52.5}{1.4} & \valstd{81.5}{0.5} & 71.1 & \gsep & \valstd{74.2}{1.4} & \valstd{77.2}{1.2} & \valstd{52.9}{2.4} & \valstd{82.2}{0.9} & 71.6 \\
 & CoreSet & \valstd{71.5}{1.8} & \valstd{77.1}{1.0} & \valstd{55.0}{1.2} & \valstd{81.8}{0.2} & 71.4 & \gsep & \valstd{70.0}{0.5} & \valstd{74.8}{0.6} & \valstd{54.7}{3.7} & \valstd{81.8}{0.9} & 70.3 \\
 & ORIENT & \valstd{72.6}{2.1} & \valstd{77.0}{0.3} & \valstd{53.4}{3.8} & \valstd{81.5}{0.8} & 71.1 & \gsep & \valstd{73.5}{2.0} & \valstd{77.2}{1.1} & \valstd{52.4}{1.7} & \valstd{81.1}{0.3} & 71.1 \\
\addlinespace[1pt]
\multirow{3}{*}{CLUE} & Random & \valstd{72.8}{2.0} & \valstd{77.2}{1.0} & \valstd{53.8}{2.2} & \valstd{82.0}{0.5} & 71.5 & \gsep & \valstd{74.3}{2.2} & \valstd{74.8}{0.4} & \valstd{56.3}{1.5} & \valstd{82.5}{0.9} & 72.0 \\
 & CoreSet & \valstd{72.3}{1.1} & \valstd{77.4}{0.3} & \valstd{55.0}{3.2} & \valstd{79.9}{0.7} & 71.2 & \gsep & \valstd{75.6}{1.1} & \valstd{77.3}{0.3} & \valstd{57.4}{2.2} & \valstd{81.8}{0.8} & 73.0 \\
 & ORIENT & \valstd{71.4}{0.6} & \valstd{77.1}{1.2} & \valstd{53.1}{2.1} & \valstd{82.4}{0.3} & 71.0 & \gsep & \valstd{74.2}{0.6} & \valstd{77.9}{1.3} & \valstd{57.2}{1.9} & \valstd{82.9}{0.4} & 73.1 \\
\addlinespace[1pt]
\multirow{3}{*}{LADA} & Random & \valstd{73.5}{0.6} & \valstd{76.5}{0.9} & \valstd{54.3}{2.9} & \valstd{81.6}{0.1} & 71.5 & \gsep & \valstd{74.3}{1.7} & \valstd{77.3}{1.1} & \valstd{54.1}{2.9} & \valstd{82.0}{0.6} & 71.9 \\
 & CoreSet & \valstd{73.0}{2.0} & \valstd{76.8}{0.5} & \valstd{52.8}{2.2} & \valstd{81.4}{0.7} & 71.0 & \gsep & \valstd{71.0}{1.5} & \valstd{74.0}{0.8} & \valstd{55.3}{2.4} & \valstd{81.7}{0.9} & 70.5 \\
 & ORIENT & \valstd{73.2}{1.0} & \valstd{77.9}{0.9} & \valstd{55.5}{2.0} & \valstd{81.6}{0.5} & 72.1 & \gsep & \valstd{74.4}{0.5} & \valstd{77.0}{0.4} & \valstd{54.4}{1.4} & \valstd{82.4}{0.2} & 72.1 \\
\textbf{TAC} & \textbf{TAC} & \valstd{74.1}{0.6} & \valstd{76.8}{0.6} & \valstd{56.2}{1.1} & \valstd{82.3}{0.5} & 72.4 & \gsep & \valstd{74.9}{1.7} & \valstd{78.1}{0.9} & \valstd{58.4}{1.1} & \valstd{82.6}{0.4} & 73.5 \\
\bottomrule
\end{tabularx}
\vspace{2pt}
\begin{tabularx}{\textwidth}{@{}l@{\hspace{3pt}\gsep\hspace{3pt}}l*{5}{Y}@{\hspace{3pt}}c@{\hspace{3pt}}*{5}{Y}@{}}
\toprule
\multirow{2}{*}{\shortstack[l]{Target\\acq.}} & \multirow{2}{*}{\shortstack[l]{Source\\sel.}} & \multicolumn{5}{c}{$B_s=200,\ B_t=10$} & \gsep & \multicolumn{5}{c}{$B_s=150,\ B_t=2$} \\
\cmidrule(lr){3-7}\cmidrule(lr){9-13}
 & & C4 & C5 & LGG & GBM & Avg. & \gsep & C4 & C5 & LGG & GBM & Avg. \\
\midrule
\multirow{3}{*}{Random} & Random & \valstd{74.1}{1.7} & \valstd{77.7}{0.8} & \valstd{60.3}{1.7} & \valstd{80.9}{0.6} & 73.3 & \gsep & \valstd{75.3}{1.0} & \valstd{78.4}{0.9} & \valstd{55.6}{3.9} & \valstd{82.4}{0.3} & 72.9 \\
 & CoreSet & \valstd{76.6}{1.3} & \valstd{76.3}{0.7} & \valstd{58.8}{2.3} & \valstd{83.7}{0.1} & 73.9 & \gsep & \valstd{74.0}{0.5} & \valstd{78.3}{1.3} & \valstd{54.0}{3.9} & \valstd{82.1}{0.5} & 72.1 \\
 & ORIENT & \valstd{75.4}{1.7} & \valstd{77.5}{0.5} & \valstd{61.8}{3.8} & \valstd{83.6}{1.0} & 74.6 & \gsep & \valstd{73.5}{1.9} & \valstd{78.8}{0.7} & \valstd{57.4}{2.5} & \valstd{82.7}{0.9} & 73.1 \\
\addlinespace[1pt]
\multirow{3}{*}{CoreSet} & Random & \valstd{71.7}{0.5} & \valstd{77.7}{0.7} & \valstd{58.5}{2.9} & \valstd{82.6}{0.3} & 72.6 & \gsep & \valstd{75.0}{1.6} & \valstd{77.9}{0.4} & \valstd{49.9}{3.5} & \valstd{82.4}{0.9} & 71.3 \\
 & CoreSet & \valstd{73.6}{0.5} & \valstd{77.6}{1.0} & \valstd{57.6}{2.2} & \valstd{82.7}{0.7} & 72.9 & \gsep & \valstd{75.2}{1.8} & \valstd{78.4}{1.3} & \valstd{52.0}{3.4} & \valstd{82.5}{0.4} & 72.0 \\
 & ORIENT & \valstd{77.3}{1.9} & \valstd{78.1}{0.4} & \valstd{62.3}{3.8} & \valstd{83.8}{0.5} & 75.4 & \gsep & \valstd{74.4}{0.9} & \valstd{77.6}{0.5} & \valstd{55.7}{2.0} & \valstd{82.3}{0.8} & 72.5 \\
\addlinespace[1pt]
\multirow{3}{*}{AADA} & Random & \valstd{74.8}{0.6} & \valstd{77.3}{0.5} & \valstd{61.3}{1.4} & \valstd{82.1}{0.2} & 73.9 & \gsep & \valstd{74.2}{1.9} & \valstd{78.1}{0.8} & \valstd{52.9}{1.5} & \valstd{81.5}{0.9} & 71.7 \\
 & CoreSet & \valstd{75.6}{1.3} & \valstd{77.0}{0.2} & \valstd{64.1}{1.9} & \valstd{83.0}{0.6} & 74.9 & \gsep & \valstd{74.5}{1.8} & \valstd{78.5}{1.2} & \valstd{53.3}{3.1} & \valstd{82.3}{0.3} & 72.2 \\
 & ORIENT & \valstd{76.1}{1.6} & \valstd{77.2}{1.2} & \valstd{61.8}{2.4} & \valstd{83.4}{0.9} & 74.6 & \gsep & \valstd{74.1}{2.2} & \valstd{78.4}{0.8} & \valstd{55.3}{3.0} & \valstd{82.2}{0.2} & 72.5 \\
\addlinespace[1pt]
\multirow{3}{*}{CLUE} & Random & \valstd{75.0}{0.6} & \valstd{77.3}{0.2} & \valstd{60.8}{1.6} & \valstd{82.5}{0.7} & 73.9 & \gsep & \valstd{73.8}{2.4} & \valstd{77.5}{0.9} & \valstd{54.9}{1.6} & \valstd{82.2}{0.5} & 72.1 \\
 & CoreSet & \valstd{77.3}{0.7} & \valstd{78.2}{0.3} & \valstd{63.7}{2.3} & \valstd{83.4}{0.6} & 75.7 & \gsep & \valstd{74.4}{2.4} & \valstd{78.1}{0.5} & \valstd{53.9}{2.1} & \valstd{81.8}{0.8} & 72.1 \\
 & ORIENT & \valstd{77.9}{1.2} & \valstd{77.7}{0.8} & \valstd{61.9}{1.8} & \valstd{81.3}{0.9} & 74.7 & \gsep & \valstd{74.1}{2.2} & \valstd{77.3}{1.0} & \valstd{56.1}{2.7} & \valstd{81.1}{0.4} & 72.2 \\
\addlinespace[1pt]
\multirow{3}{*}{LADA} & Random & \valstd{76.4}{0.5} & \valstd{77.1}{0.5} & \valstd{60.8}{1.7} & \valstd{82.1}{0.5} & 74.1 & \gsep & \valstd{75.3}{0.8} & \valstd{77.3}{0.4} & \valstd{57.8}{1.8} & \valstd{82.5}{0.5} & 73.2 \\
 & CoreSet & \valstd{77.9}{2.3} & \valstd{77.4}{1.2} & \valstd{65.3}{1.2} & \valstd{83.1}{0.8} & 75.9 & \gsep & \valstd{74.8}{0.8} & \valstd{78.2}{1.1} & \valstd{57.3}{2.0} & \valstd{82.1}{0.6} & 73.1 \\
 & ORIENT & \valstd{73.9}{0.5} & \valstd{76.3}{1.2} & \valstd{61.2}{1.7} & \valstd{83.2}{0.6} & 73.7 & \gsep & \valstd{74.4}{1.7} & \valstd{77.0}{0.9} & \valstd{55.9}{3.7} & \valstd{82.7}{0.4} & 72.5 \\
\textbf{TAC} & \textbf{TAC} & \valstd{78.5}{2.0} & \valstd{79.2}{1.2} & \valstd{64.9}{0.9} & \valstd{83.4}{1.0} & 76.5 & \gsep & \valstd{75.5}{2.1} & \valstd{77.6}{0.3} & \valstd{55.7}{3.7} & \valstd{82.3}{0.2} & 72.8 \\
\bottomrule
\end{tabularx}
\vspace{2pt}
\begin{tabularx}{\textwidth}{@{}l@{\hspace{3pt}\gsep\hspace{3pt}}l*{5}{Y}@{\hspace{3pt}}c@{\hspace{3pt}}*{5}{Y}@{}}
\toprule
\multirow{2}{*}{\shortstack[l]{Target\\acq.}} & \multirow{2}{*}{\shortstack[l]{Source\\sel.}} & \multicolumn{5}{c}{$B_s=150,\ B_t=5$} & \gsep & \multicolumn{5}{c}{$B_s=150,\ B_t=8$} \\
\cmidrule(lr){3-7}\cmidrule(lr){9-13}
 & & C4 & C5 & LGG & GBM & Avg. & \gsep & C4 & C5 & LGG & GBM & Avg. \\
\midrule
\multirow{3}{*}{Random} & Random & \valstd{75.7}{1.7} & \valstd{76.6}{0.6} & \valstd{56.9}{2.0} & \valstd{81.3}{0.1} & 72.6 & \gsep & \valstd{73.8}{2.3} & \valstd{77.6}{0.2} & \valstd{58.5}{1.4} & \valstd{82.7}{0.1} & 73.2 \\
 & CoreSet & \valstd{75.6}{0.7} & \valstd{78.2}{1.3} & \valstd{59.4}{1.3} & \valstd{82.0}{0.9} & 73.8 & \gsep & \valstd{74.8}{0.8} & \valstd{77.4}{1.0} & \valstd{62.0}{1.8} & \valstd{82.2}{0.8} & 74.1 \\
 & ORIENT & \valstd{74.9}{0.9} & \valstd{77.7}{1.3} & \valstd{61.5}{3.9} & \valstd{81.8}{0.7} & 74.0 & \gsep & \valstd{74.2}{1.0} & \valstd{76.8}{0.6} & \valstd{61.1}{2.9} & \valstd{83.1}{0.6} & 73.8 \\
\addlinespace[1pt]
\multirow{3}{*}{CoreSet} & Random & \valstd{73.9}{1.4} & \valstd{75.9}{1.2} & \valstd{58.0}{2.8} & \valstd{82.6}{0.9} & 72.6 & \gsep & \valstd{74.9}{0.5} & \valstd{77.3}{0.8} & \valstd{56.9}{3.8} & \valstd{83.2}{0.1} & 73.1 \\
 & CoreSet & \valstd{74.0}{1.8} & \valstd{77.5}{0.5} & \valstd{62.0}{3.5} & \valstd{82.3}{0.7} & 74.0 & \gsep & \valstd{78.9}{0.7} & \valstd{77.6}{1.2} & \valstd{59.4}{2.1} & \valstd{82.0}{0.2} & 74.5 \\
 & ORIENT & \valstd{73.5}{1.3} & \valstd{77.3}{0.3} & \valstd{60.6}{1.6} & \valstd{81.0}{0.8} & 73.1 & \gsep & \valstd{77.8}{0.6} & \valstd{78.2}{0.8} & \valstd{62.7}{2.6} & \valstd{82.1}{0.5} & 75.2 \\
\addlinespace[1pt]
\multirow{3}{*}{AADA} & Random & \valstd{73.2}{2.0} & \valstd{77.4}{1.1} & \valstd{52.8}{1.8} & \valstd{81.3}{0.2} & 71.2 & \gsep & \valstd{73.7}{2.0} & \valstd{77.6}{1.2} & \valstd{58.4}{1.7} & \valstd{82.8}{0.7} & 73.1 \\
 & CoreSet & \valstd{77.3}{1.4} & \valstd{77.6}{1.3} & \valstd{61.6}{2.3} & \valstd{82.8}{0.8} & 74.8 & \gsep & \valstd{78.0}{2.2} & \valstd{78.1}{0.3} & \valstd{62.7}{2.8} & \valstd{82.7}{0.6} & 75.4 \\
 & ORIENT & \valstd{74.2}{0.4} & \valstd{77.0}{0.5} & \valstd{63.1}{3.8} & \valstd{82.9}{0.6} & 74.3 & \gsep & \valstd{77.7}{1.5} & \valstd{77.9}{0.5} & \valstd{63.4}{1.2} & \valstd{82.4}{0.9} & 75.4 \\
\addlinespace[1pt]
\multirow{3}{*}{CLUE} & Random & \valstd{73.1}{2.4} & \valstd{75.7}{0.8} & \valstd{55.1}{3.9} & \valstd{82.2}{0.2} & 71.5 & \gsep & \valstd{75.0}{0.7} & \valstd{77.2}{0.8} & \valstd{62.3}{2.2} & \valstd{82.5}{0.2} & 74.3 \\
 & CoreSet & \valstd{73.6}{0.7} & \valstd{77.0}{0.7} & \valstd{64.5}{2.3} & \valstd{82.5}{0.2} & 74.4 & \gsep & \valstd{78.4}{0.7} & \valstd{78.1}{0.7} & \valstd{60.7}{2.5} & \valstd{82.4}{0.3} & 74.9 \\
 & ORIENT & \valstd{75.1}{2.0} & \valstd{78.3}{0.3} & \valstd{60.5}{2.1} & \valstd{82.7}{0.7} & 74.2 & \gsep & \valstd{74.9}{0.4} & \valstd{77.7}{0.6} & \valstd{61.5}{1.8} & \valstd{81.8}{0.3} & 74.0 \\
\addlinespace[1pt]
\multirow{3}{*}{LADA} & Random & \valstd{74.5}{2.2} & \valstd{75.3}{0.4} & \valstd{54.8}{3.9} & \valstd{81.9}{0.9} & 71.6 & \gsep & \valstd{74.3}{0.9} & \valstd{78.1}{1.1} & \valstd{56.0}{2.4} & \valstd{83.0}{0.8} & 72.9 \\
 & CoreSet & \valstd{73.7}{1.6} & \valstd{77.0}{0.6} & \valstd{62.3}{4.2} & \valstd{82.4}{0.7} & 73.9 & \gsep & \valstd{74.8}{1.0} & \valstd{77.7}{0.8} & \valstd{60.2}{2.3} & \valstd{82.7}{0.4} & 73.9 \\
 & ORIENT & \valstd{75.2}{2.3} & \valstd{78.3}{1.3} & \valstd{62.4}{1.8} & \valstd{82.0}{0.9} & 74.5 & \gsep & \valstd{74.9}{1.8} & \valstd{77.9}{0.2} & \valstd{62.9}{1.6} & \valstd{82.2}{0.5} & 74.5 \\
\textbf{TAC} & \textbf{TAC} & \valstd{78.5}{2.2} & \valstd{77.8}{1.2} & \valstd{59.4}{2.1} & \valstd{81.8}{0.5} & 74.4 & \gsep & \valstd{75.6}{2.3} & \valstd{77.2}{0.4} & \valstd{67.1}{1.6} & \valstd{82.3}{0.5} & 75.6 \\
\bottomrule
\end{tabularx}
\vspace{2pt}
\begin{tabularx}{\textwidth}{@{}l@{\hspace{3pt}\gsep\hspace{3pt}}l*{5}{Y}@{\hspace{3pt}}c@{\hspace{3pt}}*{5}{Y}@{}}
\toprule
\multirow{2}{*}{\shortstack[l]{Target\\acq.}} & \multirow{2}{*}{\shortstack[l]{Source\\sel.}} & \multicolumn{5}{c}{$B_s=150,\ B_t=10$} & \gsep & \multicolumn{5}{c}{$B_s=150,\ B_t=15$} \\
\cmidrule(lr){3-7}\cmidrule(lr){9-13}
 & & C4 & C5 & LGG & GBM & Avg. & \gsep & C4 & C5 & LGG & GBM & Avg. \\
\midrule
\multirow{3}{*}{Random} & Random & \valstd{73.9}{1.3} & \valstd{77.6}{0.4} & \valstd{60.0}{2.1} & \valstd{82.9}{0.2} & 73.6 & \gsep & \valstd{74.8}{1.8} & \valstd{77.1}{1.2} & \valstd{59.6}{1.8} & \valstd{82.4}{0.1} & 73.5 \\
 & CoreSet & \valstd{74.3}{0.1} & \valstd{77.6}{1.1} & \valstd{62.1}{1.0} & \valstd{82.8}{0.6} & 74.2 & \gsep & \valstd{73.9}{0.8} & \valstd{77.6}{0.3} & \valstd{61.2}{1.9} & \valstd{81.9}{0.5} & 73.7 \\
 & ORIENT & \valstd{75.4}{0.6} & \valstd{77.6}{0.2} & \valstd{63.7}{1.7} & \valstd{83.7}{0.3} & 75.1 & \gsep & \valstd{75.6}{2.2} & \valstd{76.8}{0.9} & \valstd{62.7}{2.8} & \valstd{83.1}{0.7} & 74.6 \\
\addlinespace[1pt]
\multirow{3}{*}{CoreSet} & Random & \valstd{74.9}{0.1} & \valstd{77.9}{0.2} & \valstd{58.7}{2.6} & \valstd{82.9}{0.1} & 73.6 & \gsep & \valstd{74.6}{0.8} & \valstd{77.3}{1.2} & \valstd{59.4}{3.0} & \valstd{82.6}{0.8} & 73.5 \\
 & CoreSet & \valstd{75.3}{0.3} & \valstd{78.3}{0.7} & \valstd{61.5}{0.2} & \valstd{82.9}{0.4} & 74.5 & \gsep & \valstd{75.8}{2.0} & \valstd{78.1}{0.4} & \valstd{60.9}{1.4} & \valstd{82.8}{0.4} & 74.4 \\
 & ORIENT & \valstd{77.8}{0.9} & \valstd{78.1}{0.6} & \valstd{61.3}{1.0} & \valstd{82.6}{0.5} & 75.0 & \gsep & \valstd{76.9}{1.8} & \valstd{77.4}{0.4} & \valstd{61.8}{2.6} & \valstd{83.2}{0.9} & 74.8 \\
\addlinespace[1pt]
\multirow{3}{*}{AADA} & Random & \valstd{76.4}{1.6} & \valstd{77.0}{0.3} & \valstd{59.8}{1.3} & \valstd{82.8}{0.1} & 74.0 & \gsep & \valstd{74.9}{1.2} & \valstd{76.6}{0.7} & \valstd{60.3}{1.8} & \valstd{82.2}{0.8} & 73.5 \\
 & CoreSet & \valstd{75.7}{0.1} & \valstd{77.8}{0.9} & \valstd{63.9}{1.8} & \valstd{82.8}{0.2} & 75.1 & \gsep & \valstd{76.4}{0.7} & \valstd{77.9}{0.7} & \valstd{62.5}{2.1} & \valstd{83.0}{0.7} & 75.0 \\
 & ORIENT & \valstd{77.4}{2.9} & \valstd{78.4}{0.5} & \valstd{63.0}{1.9} & \valstd{83.0}{0.2} & 75.5 & \gsep & \valstd{77.8}{0.9} & \valstd{78.6}{1.2} & \valstd{64.1}{2.4} & \valstd{83.4}{0.5} & 76.0 \\
\addlinespace[1pt]
\multirow{3}{*}{CLUE} & Random & \valstd{74.5}{0.8} & \valstd{76.9}{0.4} & \valstd{61.6}{1.7} & \valstd{83.1}{0.1} & 74.0 & \gsep & \valstd{74.6}{1.3} & \valstd{77.2}{1.0} & \valstd{59.9}{1.3} & \valstd{82.7}{0.7} & 73.6 \\
 & CoreSet & \valstd{75.6}{1.0} & \valstd{77.8}{0.1} & \valstd{63.0}{0.7} & \valstd{82.9}{0.1} & 74.8 & \gsep & \valstd{76.2}{1.4} & \valstd{78.3}{1.1} & \valstd{63.1}{2.6} & \valstd{82.5}{0.6} & 75.0 \\
 & ORIENT & \valstd{77.4}{0.7} & \valstd{77.6}{0.8} & \valstd{64.4}{0.3} & \valstd{82.6}{0.5} & 75.5 & \gsep & \valstd{76.7}{2.3} & \valstd{77.9}{0.3} & \valstd{62.8}{3.4} & \valstd{83.3}{0.5} & 75.2 \\
\addlinespace[1pt]
\multirow{3}{*}{LADA} & Random & \valstd{74.6}{0.9} & \valstd{77.5}{0.1} & \valstd{62.3}{3.3} & \valstd{82.6}{0.2} & 74.2 & \gsep & \valstd{74.8}{1.0} & \valstd{77.0}{0.5} & \valstd{59.6}{2.0} & \valstd{82.3}{0.2} & 73.4 \\
 & CoreSet & \valstd{75.6}{0.2} & \valstd{77.1}{1.1} & \valstd{65.9}{1.7} & \valstd{83.0}{0.1} & 75.4 & \gsep & \valstd{76.8}{0.8} & \valstd{78.2}{1.3} & \valstd{64.4}{3.1} & \valstd{83.1}{0.4} & 75.6 \\
 & ORIENT & \valstd{74.1}{0.5} & \valstd{77.8}{0.2} & \valstd{63.9}{0.1} & \valstd{83.3}{0.1} & 74.8 & \gsep & \valstd{76.1}{2.0} & \valstd{77.6}{0.8} & \valstd{63.5}{3.5} & \valstd{82.6}{0.2} & 74.9 \\
\textbf{TAC} & \textbf{TAC} & \valstd{79.1}{0.3} & \valstd{78.9}{0.3} & \valstd{67.6}{0.8} & \valstd{83.6}{0.2} & 77.3 & \gsep & \valstd{79.4}{0.5} & \valstd{78.9}{0.2} & \valstd{67.9}{1.0} & \valstd{83.5}{0.5} & 77.4 \\
\bottomrule
\end{tabularx}
\end{table*}

\begin{table*}[p]
\centering
\caption{\textbf{MAMA-MIA complete budget sweeps} (Dice \%). Source sweeps fix $B_t=10$ and target sweeps fix $B_s=150$. Entries are mean $\pm$ sample std over three seeds. Avg.: unweighted four-center mean.}
\label{tab:budget_sweep_mamamia_full}
\fontsize{6.5pt}{7.4pt}\selectfont
\setlength{\tabcolsep}{0.9pt}
\renewcommand{\arraystretch}{0.96}
\begin{tabularx}{\textwidth}{@{}l@{\hspace{3pt}\gsep\hspace{3pt}}l*{5}{Y}@{\hspace{3pt}}c@{\hspace{3pt}}*{5}{Y}@{}}
\toprule
\multirow{2}{*}{\shortstack[l]{Target\\acq.}} & \multirow{2}{*}{\shortstack[l]{Source\\sel.}} & \multicolumn{5}{c}{$B_s=50,\ B_t=10$} & \gsep & \multicolumn{5}{c}{$B_s=100,\ B_t=10$} \\
\cmidrule(lr){3-7}\cmidrule(lr){9-13}
 & & NACT & I-SPY1 & Duke & I-SPY2 & Avg. & \gsep & NACT & I-SPY1 & Duke & I-SPY2 & Avg. \\
\midrule
\multirow{3}{*}{Random} & Random & \valstd{66.1}{1.7} & \valstd{64.0}{1.1} & \valstd{48.5}{2.0} & \valstd{60.5}{0.4} & 59.7 & \gsep & \valstd{66.8}{1.5} & \valstd{66.8}{1.6} & \valstd{55.7}{2.1} & \valstd{65.1}{1.2} & 63.6 \\
 & CoreSet & \valstd{67.1}{0.4} & \valstd{62.1}{0.9} & \valstd{53.8}{1.3} & \valstd{58.0}{0.8} & 60.3 & \gsep & \valstd{69.4}{1.4} & \valstd{68.0}{1.4} & \valstd{53.1}{2.4} & \valstd{63.4}{1.2} & 63.5 \\
 & ORIENT & \valstd{67.4}{0.9} & \valstd{65.8}{0.4} & \valstd{54.9}{2.2} & \valstd{57.1}{1.1} & 61.3 & \gsep & \valstd{63.7}{1.6} & \valstd{67.9}{0.3} & \valstd{54.3}{1.4} & \valstd{64.9}{0.5} & 62.7 \\
\addlinespace[1pt]
\multirow{3}{*}{CoreSet} & Random & \valstd{59.7}{1.1} & \valstd{64.2}{1.4} & \valstd{54.1}{1.5} & \valstd{64.8}{0.3} & 60.7 & \gsep & \valstd{65.5}{1.9} & \valstd{69.3}{1.4} & \valstd{58.3}{3.2} & \valstd{67.3}{0.2} & 65.1 \\
 & CoreSet & \valstd{67.4}{0.8} & \valstd{62.2}{0.4} & \valstd{52.6}{1.5} & \valstd{60.7}{0.3} & 60.7 & \gsep & \valstd{69.3}{0.5} & \valstd{65.3}{0.5} & \valstd{55.3}{2.0} & \valstd{63.7}{0.4} & 63.4 \\
 & ORIENT & \valstd{66.6}{2.1} & \valstd{61.7}{0.6} & \valstd{51.8}{2.8} & \valstd{59.2}{0.4} & 59.8 & \gsep & \valstd{68.3}{0.8} & \valstd{63.5}{1.4} & \valstd{53.5}{1.9} & \valstd{65.0}{0.5} & 62.6 \\
\addlinespace[1pt]
\multirow{3}{*}{AADA} & Random & \valstd{61.1}{0.9} & \valstd{64.9}{0.3} & \valstd{53.5}{1.3} & \valstd{63.3}{0.7} & 60.7 & \gsep & \valstd{64.3}{1.2} & \valstd{69.5}{1.0} & \valstd{56.7}{1.8} & \valstd{65.7}{0.4} & 64.0 \\
 & CoreSet & \valstd{66.5}{1.3} & \valstd{62.3}{0.7} & \valstd{45.7}{1.8} & \valstd{57.9}{1.4} & 58.1 & \gsep & \valstd{68.4}{1.4} & \valstd{67.5}{0.2} & \valstd{51.3}{2.6} & \valstd{63.7}{1.4} & 62.7 \\
 & ORIENT & \valstd{66.4}{1.0} & \valstd{59.6}{1.6} & \valstd{46.5}{1.0} & \valstd{60.9}{1.3} & 58.4 & \gsep & \valstd{66.9}{0.6} & \valstd{67.1}{1.4} & \valstd{52.9}{2.9} & \valstd{67.1}{1.4} & 63.5 \\
\addlinespace[1pt]
\multirow{3}{*}{CLUE} & Random & \valstd{60.1}{0.8} & \valstd{66.3}{1.1} & \valstd{54.6}{1.9} & \valstd{63.5}{0.4} & 61.1 & \gsep & \valstd{69.4}{0.6} & \valstd{67.1}{1.0} & \valstd{51.4}{1.1} & \valstd{63.9}{0.3} & 63.0 \\
 & CoreSet & \valstd{69.3}{0.3} & \valstd{62.8}{0.7} & \valstd{48.7}{2.8} & \valstd{59.5}{0.8} & 60.1 & \gsep & \valstd{67.2}{1.4} & \valstd{70.0}{0.2} & \valstd{58.9}{0.9} & \valstd{66.1}{0.9} & 65.5 \\
 & ORIENT & \valstd{60.1}{0.6} & \valstd{65.6}{0.2} & \valstd{58.3}{0.8} & \valstd{63.6}{1.0} & 61.9 & \gsep & \valstd{69.4}{2.0} & \valstd{71.1}{0.9} & \valstd{56.9}{0.6} & \valstd{67.4}{0.4} & 66.2 \\
\addlinespace[1pt]
\multirow{3}{*}{LADA} & Random & \valstd{63.7}{1.5} & \valstd{65.9}{1.3} & \valstd{58.2}{1.3} & \valstd{63.5}{0.8} & 62.8 & \gsep & \valstd{66.8}{0.4} & \valstd{69.5}{1.0} & \valstd{60.3}{1.5} & \valstd{65.5}{0.8} & 65.5 \\
 & CoreSet & \valstd{65.7}{1.7} & \valstd{64.5}{0.5} & \valstd{54.8}{2.9} & \valstd{60.8}{0.8} & 61.4 & \gsep & \valstd{68.6}{1.4} & \valstd{66.7}{0.3} & \valstd{56.8}{2.1} & \valstd{65.5}{0.9} & 64.4 \\
 & ORIENT & \valstd{64.5}{0.7} & \valstd{64.8}{1.0} & \valstd{56.3}{0.7} & \valstd{62.5}{0.7} & 62.0 & \gsep & \valstd{64.6}{2.0} & \valstd{68.1}{1.1} & \valstd{57.1}{2.3} & \valstd{64.7}{1.2} & 63.6 \\
\textbf{TAC} & \textbf{TAC} & \valstd{68.4}{1.5} & \valstd{67.6}{0.9} & \valstd{59.7}{1.9} & \valstd{62.3}{1.2} & 64.5 & \gsep & \valstd{69.5}{1.3} & \valstd{68.1}{0.9} & \valstd{61.4}{1.9} & \valstd{66.6}{0.4} & 66.4 \\
\bottomrule
\end{tabularx}
\vspace{2pt}
\begin{tabularx}{\textwidth}{@{}l@{\hspace{3pt}\gsep\hspace{3pt}}l*{5}{Y}@{\hspace{3pt}}c@{\hspace{3pt}}*{5}{Y}@{}}
\toprule
\multirow{2}{*}{\shortstack[l]{Target\\acq.}} & \multirow{2}{*}{\shortstack[l]{Source\\sel.}} & \multicolumn{5}{c}{$B_s=200,\ B_t=10$} & \gsep & \multicolumn{5}{c}{$B_s=150,\ B_t=2$} \\
\cmidrule(lr){3-7}\cmidrule(lr){9-13}
 & & NACT & I-SPY1 & Duke & I-SPY2 & Avg. & \gsep & NACT & I-SPY1 & Duke & I-SPY2 & Avg. \\
\midrule
\multirow{3}{*}{Random} & Random & \valstd{67.0}{0.6} & \valstd{66.9}{1.0} & \valstd{53.1}{2.4} & \valstd{67.6}{1.1} & 63.7 & \gsep & \valstd{61.3}{1.1} & \valstd{63.8}{0.7} & \valstd{51.0}{2.5} & \valstd{64.6}{0.5} & 60.2 \\
 & CoreSet & \valstd{66.0}{2.0} & \valstd{68.2}{1.1} & \valstd{56.8}{1.6} & \valstd{65.3}{0.8} & 64.1 & \gsep & \valstd{63.9}{1.8} & \valstd{63.1}{1.5} & \valstd{49.6}{3.0} & \valstd{64.2}{1.3} & 60.2 \\
 & ORIENT & \valstd{65.9}{1.8} & \valstd{67.2}{0.7} & \valstd{58.7}{2.0} & \valstd{65.0}{0.8} & 64.2 & \gsep & \valstd{63.6}{0.4} & \valstd{63.3}{1.0} & \valstd{42.9}{2.5} & \valstd{63.3}{1.1} & 58.3 \\
\addlinespace[1pt]
\multirow{3}{*}{CoreSet} & Random & \valstd{66.3}{0.9} & \valstd{66.3}{1.6} & \valstd{61.3}{3.1} & \valstd{67.4}{0.9} & 65.3 & \gsep & \valstd{60.6}{0.7} & \valstd{65.1}{1.5} & \valstd{59.0}{2.9} & \valstd{67.4}{1.0} & 63.0 \\
 & CoreSet & \valstd{68.4}{0.6} & \valstd{67.4}{0.5} & \valstd{56.6}{1.3} & \valstd{66.5}{1.3} & 64.7 & \gsep & \valstd{63.4}{1.1} & \valstd{63.6}{1.2} & \valstd{54.7}{3.0} & \valstd{62.5}{1.2} & 61.1 \\
 & ORIENT & \valstd{68.3}{0.8} & \valstd{66.4}{0.5} & \valstd{55.4}{2.0} & \valstd{67.8}{0.4} & 64.5 & \gsep & \valstd{62.3}{2.1} & \valstd{65.4}{0.8} & \valstd{50.8}{0.7} & \valstd{64.5}{0.6} & 60.7 \\
\addlinespace[1pt]
\multirow{3}{*}{AADA} & Random & \valstd{69.5}{1.7} & \valstd{67.1}{0.8} & \valstd{58.4}{1.1} & \valstd{69.0}{0.3} & 66.0 & \gsep & \valstd{64.9}{0.4} & \valstd{66.8}{1.0} & \valstd{58.5}{1.8} & \valstd{68.4}{1.0} & 64.6 \\
 & CoreSet & \valstd{67.5}{0.5} & \valstd{68.8}{0.6} & \valstd{56.1}{2.5} & \valstd{68.6}{0.6} & 65.2 & \gsep & \valstd{67.8}{1.5} & \valstd{67.3}{0.6} & \valstd{49.5}{3.2} & \valstd{63.2}{1.0} & 61.9 \\
 & ORIENT & \valstd{67.1}{1.0} & \valstd{68.3}{1.2} & \valstd{56.1}{1.8} & \valstd{68.7}{1.3} & 65.0 & \gsep & \valstd{60.3}{1.6} & \valstd{64.4}{0.6} & \valstd{48.0}{3.1} & \valstd{63.1}{1.0} & 59.0 \\
\addlinespace[1pt]
\multirow{3}{*}{CLUE} & Random & \valstd{67.6}{1.9} & \valstd{70.0}{1.2} & \valstd{60.0}{1.8} & \valstd{66.4}{0.9} & 66.0 & \gsep & \valstd{61.8}{0.9} & \valstd{66.3}{1.2} & \valstd{60.5}{1.6} & \valstd{68.5}{0.8} & 64.3 \\
 & CoreSet & \valstd{67.5}{1.6} & \valstd{67.9}{1.0} & \valstd{56.3}{1.7} & \valstd{66.0}{0.4} & 64.5 & \gsep & \valstd{61.3}{1.4} & \valstd{63.5}{0.4} & \valstd{56.3}{2.1} & \valstd{66.0}{1.1} & 61.8 \\
 & ORIENT & \valstd{66.2}{1.0} & \valstd{67.8}{0.8} & \valstd{61.5}{2.6} & \valstd{66.2}{0.9} & 65.4 & \gsep & \valstd{65.9}{1.3} & \valstd{66.2}{0.5} & \valstd{58.5}{3.1} & \valstd{64.4}{0.6} & 63.7 \\
\addlinespace[1pt]
\multirow{3}{*}{LADA} & Random & \valstd{64.0}{1.6} & \valstd{65.6}{1.5} & \valstd{60.6}{1.4} & \valstd{66.8}{0.5} & 64.2 & \gsep & \valstd{63.0}{2.0} & \valstd{65.4}{1.1} & \valstd{57.7}{1.2} & \valstd{69.3}{0.3} & 63.9 \\
 & CoreSet & \valstd{69.5}{0.7} & \valstd{68.1}{1.0} & \valstd{56.6}{2.2} & \valstd{67.2}{0.2} & 65.4 & \gsep & \valstd{62.0}{2.0} & \valstd{66.3}{0.3} & \valstd{55.6}{3.5} & \valstd{64.8}{0.8} & 62.2 \\
 & ORIENT & \valstd{67.8}{1.7} & \valstd{67.1}{1.3} & \valstd{56.0}{3.0} & \valstd{67.4}{0.7} & 64.6 & \gsep & \valstd{61.5}{1.2} & \valstd{65.6}{0.5} & \valstd{46.3}{1.5} & \valstd{66.7}{0.2} & 60.0 \\
\textbf{TAC} & \textbf{TAC} & \valstd{71.6}{1.0} & \valstd{69.4}{1.5} & \valstd{61.7}{3.0} & \valstd{69.3}{0.3} & 68.0 & \gsep & \valstd{65.9}{0.9} & \valstd{67.4}{0.8} & \valstd{58.7}{2.4} & \valstd{65.1}{0.8} & 64.3 \\
\bottomrule
\end{tabularx}
\vspace{2pt}
\begin{tabularx}{\textwidth}{@{}l@{\hspace{3pt}\gsep\hspace{3pt}}l*{5}{Y}@{\hspace{3pt}}c@{\hspace{3pt}}*{5}{Y}@{}}
\toprule
\multirow{2}{*}{\shortstack[l]{Target\\acq.}} & \multirow{2}{*}{\shortstack[l]{Source\\sel.}} & \multicolumn{5}{c}{$B_s=150,\ B_t=5$} & \gsep & \multicolumn{5}{c}{$B_s=150,\ B_t=8$} \\
\cmidrule(lr){3-7}\cmidrule(lr){9-13}
 & & NACT & I-SPY1 & Duke & I-SPY2 & Avg. & \gsep & NACT & I-SPY1 & Duke & I-SPY2 & Avg. \\
\midrule
\multirow{3}{*}{Random} & Random & \valstd{63.3}{1.8} & \valstd{66.5}{0.5} & \valstd{49.7}{2.8} & \valstd{65.8}{1.0} & 61.3 & \gsep & \valstd{60.9}{0.3} & \valstd{65.1}{0.4} & \valstd{56.0}{1.0} & \valstd{66.9}{0.9} & 62.3 \\
 & CoreSet & \valstd{64.2}{1.8} & \valstd{67.2}{1.4} & \valstd{50.1}{3.5} & \valstd{65.2}{0.9} & 61.7 & \gsep & \valstd{60.8}{1.4} & \valstd{65.3}{1.1} & \valstd{56.5}{1.6} & \valstd{66.2}{0.8} & 62.2 \\
 & ORIENT & \valstd{55.6}{0.5} & \valstd{64.9}{0.7} & \valstd{40.5}{1.8} & \valstd{64.7}{1.3} & 56.4 & \gsep & \valstd{65.5}{0.7} & \valstd{64.9}{0.9} & \valstd{53.7}{0.8} & \valstd{66.5}{0.3} & 62.7 \\
\addlinespace[1pt]
\multirow{3}{*}{CoreSet} & Random & \valstd{68.6}{1.2} & \valstd{67.6}{1.0} & \valstd{46.6}{2.4} & \valstd{68.5}{1.4} & 62.8 & \gsep & \valstd{66.3}{1.2} & \valstd{67.6}{1.1} & \valstd{59.5}{3.6} & \valstd{69.2}{0.4} & 65.6 \\
 & CoreSet & \valstd{65.6}{1.7} & \valstd{66.0}{0.4} & \valstd{51.9}{3.6} & \valstd{67.3}{1.1} & 62.7 & \gsep & \valstd{62.1}{1.7} & \valstd{65.4}{1.4} & \valstd{54.5}{0.8} & \valstd{64.0}{0.3} & 61.5 \\
 & ORIENT & \valstd{66.0}{1.1} & \valstd{66.8}{0.3} & \valstd{53.6}{3.7} & \valstd{67.6}{0.4} & 63.5 & \gsep & \valstd{62.1}{0.7} & \valstd{64.7}{1.2} & \valstd{52.3}{1.0} & \valstd{67.8}{1.2} & 61.7 \\
\addlinespace[1pt]
\multirow{3}{*}{AADA} & Random & \valstd{57.1}{0.9} & \valstd{69.3}{0.9} & \valstd{49.6}{3.4} & \valstd{68.7}{0.3} & 61.1 & \gsep & \valstd{67.2}{0.8} & \valstd{69.0}{0.7} & \valstd{59.5}{3.0} & \valstd{67.9}{0.6} & 65.9 \\
 & CoreSet & \valstd{61.5}{1.1} & \valstd{67.1}{0.5} & \valstd{51.2}{2.6} & \valstd{67.4}{0.7} & 61.8 & \gsep & \valstd{67.9}{1.5} & \valstd{68.4}{0.2} & \valstd{53.1}{2.5} & \valstd{65.2}{1.3} & 63.6 \\
 & ORIENT & \valstd{67.9}{0.7} & \valstd{66.9}{0.5} & \valstd{55.8}{3.2} & \valstd{65.0}{0.4} & 63.9 & \gsep & \valstd{65.5}{1.2} & \valstd{67.4}{0.8} & \valstd{49.7}{1.2} & \valstd{65.3}{0.6} & 62.0 \\
\addlinespace[1pt]
\multirow{3}{*}{CLUE} & Random & \valstd{66.6}{1.7} & \valstd{68.0}{1.1} & \valstd{53.8}{1.4} & \valstd{71.0}{0.9} & 64.9 & \gsep & \valstd{65.0}{2.1} & \valstd{67.6}{0.4} & \valstd{61.1}{3.1} & \valstd{68.9}{0.4} & 65.7 \\
 & CoreSet & \valstd{66.4}{0.8} & \valstd{64.4}{0.4} & \valstd{53.1}{2.6} & \valstd{67.1}{1.0} & 62.7 & \gsep & \valstd{67.0}{1.6} & \valstd{65.7}{1.0} & \valstd{56.0}{2.6} & \valstd{66.0}{0.4} & 63.7 \\
 & ORIENT & \valstd{64.7}{0.8} & \valstd{68.7}{0.4} & \valstd{49.7}{3.9} & \valstd{67.1}{0.3} & 62.5 & \gsep & \valstd{61.6}{1.8} & \valstd{68.0}{0.7} & \valstd{61.2}{3.5} & \valstd{69.2}{0.7} & 65.0 \\
\addlinespace[1pt]
\multirow{3}{*}{LADA} & Random & \valstd{64.8}{2.0} & \valstd{69.0}{1.2} & \valstd{46.3}{3.2} & \valstd{70.2}{0.3} & 62.6 & \gsep & \valstd{61.3}{1.5} & \valstd{67.6}{0.2} & \valstd{58.9}{2.7} & \valstd{68.9}{0.7} & 64.2 \\
 & CoreSet & \valstd{64.1}{0.6} & \valstd{64.4}{0.9} & \valstd{55.5}{1.7} & \valstd{67.8}{0.8} & 63.0 & \gsep & \valstd{64.6}{1.4} & \valstd{67.0}{0.4} & \valstd{50.3}{2.2} & \valstd{67.0}{0.7} & 62.2 \\
 & ORIENT & \valstd{69.0}{1.9} & \valstd{63.6}{1.5} & \valstd{47.1}{1.5} & \valstd{67.5}{0.6} & 61.8 & \gsep & \valstd{64.1}{1.1} & \valstd{66.5}{0.6} & \valstd{44.5}{3.8} & \valstd{63.6}{1.2} & 59.7 \\
\textbf{TAC} & \textbf{TAC} & \valstd{68.5}{0.6} & \valstd{68.1}{1.4} & \valstd{56.7}{3.5} & \valstd{68.5}{1.2} & 65.5 & \gsep & \valstd{70.0}{1.0} & \valstd{67.4}{0.2} & \valstd{59.1}{3.0} & \valstd{68.4}{0.4} & 66.2 \\
\bottomrule
\end{tabularx}
\vspace{2pt}
\begin{tabularx}{\textwidth}{@{}l@{\hspace{3pt}\gsep\hspace{3pt}}l*{5}{Y}@{\hspace{3pt}}c@{\hspace{3pt}}*{5}{Y}@{}}
\toprule
\multirow{2}{*}{\shortstack[l]{Target\\acq.}} & \multirow{2}{*}{\shortstack[l]{Source\\sel.}} & \multicolumn{5}{c}{$B_s=150,\ B_t=10$} & \gsep & \multicolumn{5}{c}{$B_s=150,\ B_t=15$} \\
\cmidrule(lr){3-7}\cmidrule(lr){9-13}
 & & NACT & I-SPY1 & Duke & I-SPY2 & Avg. & \gsep & NACT & I-SPY1 & Duke & I-SPY2 & Avg. \\
\midrule
\multirow{3}{*}{Random} & Random & \valstd{65.2}{0.2} & \valstd{66.3}{0.4} & \valstd{56.2}{1.7} & \valstd{66.6}{1.4} & 63.6 & \gsep & \valstd{62.9}{1.6} & \valstd{67.1}{0.4} & \valstd{58.0}{2.5} & \valstd{68.9}{0.3} & 64.3 \\
 & CoreSet & \valstd{67.2}{0.5} & \valstd{67.4}{0.1} & \valstd{56.4}{0.5} & \valstd{65.7}{0.2} & 64.2 & \gsep & \valstd{62.8}{1.4} & \valstd{67.3}{0.5} & \valstd{58.5}{0.8} & \valstd{68.2}{1.2} & 64.2 \\
 & ORIENT & \valstd{62.2}{1.4} & \valstd{67.0}{0.1} & \valstd{55.3}{0.4} & \valstd{66.2}{0.5} & 62.7 & \gsep & \valstd{67.5}{1.7} & \valstd{66.9}{1.1} & \valstd{55.7}{1.8} & \valstd{68.5}{0.7} & 64.7 \\
\addlinespace[1pt]
\multirow{3}{*}{CoreSet} & Random & \valstd{68.0}{0.3} & \valstd{68.0}{0.5} & \valstd{59.8}{0.4} & \valstd{69.7}{0.2} & 66.4 & \gsep & \valstd{68.3}{1.3} & \valstd{69.6}{0.5} & \valstd{61.5}{2.9} & \valstd{71.2}{0.3} & 67.6 \\
 & CoreSet & \valstd{68.0}{0.5} & \valstd{66.1}{0.3} & \valstd{55.2}{0.6} & \valstd{66.2}{0.3} & 63.9 & \gsep & \valstd{64.1}{0.6} & \valstd{67.4}{0.6} & \valstd{56.5}{2.3} & \valstd{66.0}{0.9} & 63.5 \\
 & ORIENT & \valstd{67.0}{1.0} & \valstd{65.3}{0.4} & \valstd{54.1}{0.2} & \valstd{67.9}{0.1} & 63.6 & \gsep & \valstd{64.1}{1.6} & \valstd{66.7}{1.0} & \valstd{54.3}{2.1} & \valstd{69.8}{1.2} & 63.7 \\
\addlinespace[1pt]
\multirow{3}{*}{AADA} & Random & \valstd{64.1}{1.3} & \valstd{68.6}{0.2} & \valstd{58.4}{0.2} & \valstd{68.6}{0.9} & 64.9 & \gsep & \valstd{69.2}{1.3} & \valstd{70.0}{0.9} & \valstd{61.5}{1.3} & \valstd{69.9}{1.1} & 67.6 \\
 & CoreSet & \valstd{63.7}{1.7} & \valstd{67.1}{0.1} & \valstd{50.9}{2.7} & \valstd{66.9}{0.2} & 62.1 & \gsep & \valstd{69.9}{0.4} & \valstd{70.4}{1.3} & \valstd{55.1}{2.9} & \valstd{67.2}{1.4} & 65.6 \\
 & ORIENT & \valstd{67.3}{0.1} & \valstd{69.3}{0.3} & \valstd{55.8}{0.1} & \valstd{67.8}{0.6} & 65.0 & \gsep & \valstd{67.5}{1.8} & \valstd{69.4}{0.9} & \valstd{51.7}{2.9} & \valstd{67.3}{0.7} & 64.0 \\
\addlinespace[1pt]
\multirow{3}{*}{CLUE} & Random & \valstd{65.1}{0.5} & \valstd{68.3}{0.1} & \valstd{59.0}{0.6} & \valstd{69.5}{0.4} & 65.5 & \gsep & \valstd{67.0}{0.8} & \valstd{66.6}{1.5} & \valstd{62.1}{2.0} & \valstd{68.9}{1.1} & 66.2 \\
 & CoreSet & \valstd{66.0}{0.3} & \valstd{67.8}{0.1} & \valstd{52.8}{2.5} & \valstd{66.9}{0.1} & 63.4 & \gsep & \valstd{69.0}{0.3} & \valstd{67.7}{0.3} & \valstd{58.0}{3.5} & \valstd{68.0}{1.0} & 65.7 \\
 & ORIENT & \valstd{64.3}{1.2} & \valstd{69.6}{0.7} & \valstd{59.8}{0.3} & \valstd{68.4}{0.3} & 65.5 & \gsep & \valstd{63.6}{1.6} & \valstd{70.0}{1.1} & \valstd{63.2}{1.1} & \valstd{71.2}{0.8} & 67.0 \\
\addlinespace[1pt]
\multirow{3}{*}{LADA} & Random & \valstd{67.5}{0.9} & \valstd{68.0}{0.2} & \valstd{60.0}{0.6} & \valstd{69.1}{0.4} & 66.1 & \gsep & \valstd{63.3}{1.2} & \valstd{69.6}{1.0} & \valstd{60.9}{3.9} & \valstd{70.9}{0.7} & 66.2 \\
 & CoreSet & \valstd{66.1}{0.2} & \valstd{67.5}{0.4} & \valstd{59.5}{0.2} & \valstd{67.4}{0.3} & 65.1 & \gsep & \valstd{66.6}{1.6} & \valstd{69.0}{1.1} & \valstd{52.3}{0.9} & \valstd{69.0}{1.4} & 64.2 \\
 & ORIENT & \valstd{67.9}{0.1} & \valstd{67.5}{0.6} & \valstd{56.9}{1.1} & \valstd{66.2}{0.3} & 64.6 & \gsep & \valstd{66.1}{1.0} & \valstd{68.5}{1.5} & \valstd{46.5}{3.6} & \valstd{65.6}{0.6} & 61.7 \\
\textbf{TAC} & \textbf{TAC} & \valstd{70.5}{1.4} & \valstd{70.3}{0.4} & \valstd{62.9}{0.8} & \valstd{70.0}{0.5} & 68.4 & \gsep & \valstd{70.0}{1.4} & \valstd{69.4}{1.1} & \valstd{63.1}{3.0} & \valstd{70.4}{0.5} & 68.2 \\
\bottomrule
\end{tabularx}
\end{table*}

\subsection{Budget Sweeps and Reference Configurations}
\label{app:budget_details}
\label{app:reference_baselines}
At fixed $B_t=10$, the source-budget sweep considers
$B_s\in\{50,100,150,200\}$; at fixed $B_s=150$, the acquisition sweep
considers $B_t\in\{2,5,8,10,15\}$. For BraTS C4, $B_t=15$ exhausts the 15-case acquisition pool, making acquisition non-selective for that center. Every plotted value is the mean of
three independent training-seed results, with each seed averaged equally
across the four held-out centers. Best selection is the strongest among the
selection-only combinations shown in the plot. Best DA is the strongest of the four domain-adaptation methods
evaluated at each corresponding budget. Each gain annotation in Figure~4 of the main paper
is TAC minus the larger of the Best selection-only and Best DA means
at that budget, in Dice points. Full Source-only is a separate reference
trained with all source labels and no target labels and is excluded from
budget-matched rankings. Seed-level reference results appear in
\cref{app:reference_seed_results}.

\Needspace{9\baselineskip}
\subsection{Complete Budget-Sweep Results}

\label{app:budget_sweep_full_tables}

Tables~\ref{tab:budget_sweep_brats_full} and~\ref{tab:budget_sweep_mamamia_full} enumerate 15 acquisition--selection controls (excluding LESS) and TAC across the budget grid. Entries report three-seed means $\pm$ sample standard deviations; the primary setting $(B_s,B_t)=(150,10)$ matches the main-paper results. Values are rounded to one decimal Dice point.

\subsection{Seed-Level Reference Results}
\label{app:reference_seed_results}
Tables~\ref{tab:reference_brats_seeds} and~\ref{tab:reference_mamamia_seeds}
report individual training-seed results for the three reference configurations
in Table~2 of the main paper. Full Source-only uses all labeled source cases
and no target annotations; Random + Full Source adds ten randomly acquired
target cases; Random target-only uses ten randomly acquired target cases alone.
Each seed average weights the four target centers equally.

\begin{table}[t]
\centering
\caption{\textbf{BraTS references by seed.} Macro Dice (\%) for the three reference configurations. Avg.: unweighted four-center mean.}
\label{tab:reference_brats_seeds}
\scriptsize
\setlength{\tabcolsep}{3pt}
\renewcommand{\arraystretch}{1.08}
\begin{tabular}{@{}lrrrrr@{}}
\toprule
Seed & C4 & C5 & LGG & GBM & Avg. \\
\midrule
\multicolumn{6}{@{}l}{\emph{Full Source-only} ($B_t=0$)} \\
0 & 73.5 & 76.9 & 59.6 & 84.1 & 73.53 \\
1 & 72.9 & 78.2 & 60.8 & 82.8 & 73.68 \\
2 & 73.1 & 75.0 & 62.5 & 84.2 & 73.70 \\
\midrule
\multicolumn{6}{@{}l}{\emph{Random + Full Source} ($B_t=10$)} \\
0 & 75.7 & 78.9 & 63.3 & 84.1 & 75.50 \\
1 & 75.1 & 76.2 & 61.5 & 83.7 & 74.13 \\
2 & 74.1 & 76.9 & 62.6 & 84.0 & 74.40 \\
\midrule
\multicolumn{6}{@{}l}{\emph{Random target-only} ($B_t=10$)} \\
0 & 67.8 & 71.8 & 51.1 & 79.4 & 67.53 \\
1 & 70.3 & 73.6 & 51.3 & 75.5 & 67.68 \\
2 & 68.3 & 72.4 & 49.5 & 78.7 & 67.23 \\
\bottomrule
\end{tabular}
\end{table}

\begin{table}[t]
\centering
\caption{\textbf{MAMA-MIA references by seed.} Foreground Dice (\%) for the three reference configurations. Avg.: unweighted four-center mean.}
\label{tab:reference_mamamia_seeds}
\scriptsize
\setlength{\tabcolsep}{3pt}
\renewcommand{\arraystretch}{1.08}
\begin{tabular}{@{}lrrrrr@{}}
\toprule
Seed & NACT & I-SPY1 & Duke & I-SPY2 & Avg. \\
\midrule
\multicolumn{6}{@{}l}{\emph{Full Source-only} ($B_t=0$)} \\
0 & 68.0 & 70.0 & 53.0 & 68.4 & 64.85 \\
1 & 67.3 & 68.1 & 56.0 & 66.8 & 64.55 \\
2 & 65.8 & 64.5 & 57.1 & 67.9 & 63.83 \\
\midrule
\multicolumn{6}{@{}l}{\emph{Random + Full Source} ($B_t=10$)} \\
0 & 69.8 & 71.1 & 55.9 & 71.2 & 67.00 \\
1 & 67.2 & 69.5 & 57.7 & 68.4 & 65.70 \\
2 & 66.4 & 69.1 & 60.9 & 67.8 & 66.05 \\
\midrule
\multicolumn{6}{@{}l}{\emph{Random target-only} ($B_t=10$)} \\
0 & 48.3 & 59.6 & 52.5 & 61.5 & 55.48 \\
1 & 49.4 & 56.8 & 53.3 & 63.0 & 55.63 \\
2 & 50.6 & 57.4 & 50.9 & 62.7 & 55.40 \\
\bottomrule
\end{tabular}
\end{table}%

\subsection{Source--Target Coupling}
\label{app:coupling_details}
At $(B_s,B_t)=(150,10)$, coupling is evaluated through a matched
$2\times2$ design: Random/Random, TAC acquisition/Random source,
Random acquisition/TAC source, and TAC/TAC. For each TAC source-selection arm, the source reliability and coverage
evidence is recomputed from the support acquired in that arm, including
reference embeddings and gradients. For each independent seed, we compute $D_{11}-D_{10}-D_{01}+D_{00}$ after the four-center average,
then summarize across seeds. The seed-resolved four-arm results and paired interactions are reported
for both benchmarks in \cref{tab:coupling_brats_seeds,tab:coupling_mamamia_seeds}.
All uncertainty values are sample standard deviations across three
matched seeds.

\begin{table}[t]
    \centering
    \caption{\textbf{BraTS factorial results by seed}, $(B_s,B_t)=(150,10)$. Four-center macro Dice (\%); matched-seed interaction. Last column: three-seed mean $\pm$ sample std.}
    \label{tab:coupling_brats_seeds}
    \footnotesize
    \setlength{\tabcolsep}{3pt}
    \renewcommand{\arraystretch}{1.12}
    \begin{tabular}{@{}lrrrr@{}}
        \toprule
        Target / source & Seed 0 & Seed 1 & Seed 2 & Mean $\pm$ std \\
        \midrule
        Random / Random & 72.55 & 74.00 & 74.35 & \valstd{73.63}{0.95} \\
        TAC / Random & 71.73 & 73.87 & 73.46 & \valstd{73.02}{1.14} \\
        Random / TAC & 74.89 & 75.47 & 74.98 & \valstd{75.11}{0.31} \\
        TAC / TAC & 77.08 & 77.23 & 77.63 & \valstd{77.31}{0.28} \\
        \midrule
        Interaction $\Delta_{\mathrm{int}}$ & +3.01 & +1.90 & +3.54 & \valstd{+2.82}{0.84} \\
        \bottomrule
    \end{tabular}
\end{table}
\begin{table}[t]
    \centering
    \caption{\textbf{BraTS hybrid arms by target and seed}, $(B_s,B_t)=(150,10)$, macro Dice (\%). Per-center scores and four-center seed means.}
    \label{tab:coupling_brats_hybrid_centers}
    \scriptsize
    \setlength{\tabcolsep}{2.0pt}
    \renewcommand{\arraystretch}{1.1}
    \begin{tabular}{@{}llrrrrr@{}}
        \toprule
        Target / source & Seed & C4 & C5 & LGG & GBM & Avg. \\
        \midrule
        \multirow{3}{*}{TAC / Random} & 0 & 72.51 & 76.37 & 55.78 & 82.26 & 71.73 \\
        & 1 & 74.82 & 77.46 & 60.39 & 82.81 & 73.87 \\
        & 2 & 74.83 & 77.51 & 58.59 & 82.90 & 73.46 \\
        \midrule
        \multirow{3}{*}{Random / TAC} & 0 & 74.03 & 78.47 & 63.44 & 83.61 & 74.89 \\
        & 1 & 74.83 & 78.30 & 65.15 & 83.58 & 75.47 \\
        & 2 & 74.52 & 78.36 & 63.44 & 83.61 & 74.98 \\
        \midrule
        \multirow{3}{*}{TAC / CoreSet} & 0 & 77.67 & 78.11 & 62.03 & 80.49 & 74.58 \\
        & 1 & 77.17 & 77.56 & 61.27 & 82.53 & 74.63 \\
        & 2 & 77.06 & 77.57 & 61.23 & 82.68 & 74.64 \\
        \midrule
        \multirow{3}{*}{CoreSet / TAC} & 0 & 74.56 & 78.29 & 61.47 & 83.10 & 74.36 \\
        & 1 & 73.94 & 77.92 & 63.68 & 83.23 & 74.69 \\
        & 2 & 74.16 & 78.55 & 63.48 & 83.85 & 75.01 \\
        \midrule
        \multirow{3}{*}{TAC / ORIENT} & 0 & 74.15 & 78.17 & 64.55 & 82.96 & 74.96 \\
        & 1 & 78.20 & 78.21 & 64.13 & 83.36 & 75.98 \\
        & 2 & 77.78 & 77.05 & 62.55 & 83.73 & 75.28 \\
        \midrule
        \multirow{3}{*}{CLUE / TAC} & 0 & 74.31 & 77.66 & 62.98 & 83.40 & 74.59 \\
        & 1 & 73.96 & 77.53 & 61.60 & 83.52 & 74.15 \\
        & 2 & 74.02 & 78.13 & 63.17 & 83.61 & 74.73 \\
        \bottomrule
    \end{tabular}
\end{table}

The seed-level MAMA-MIA scores and interactions in
\cref{tab:coupling_mamamia_seeds} use the same matching across all four
factorial arms. Each seed score is the unweighted mean of its four target-center scores.
Per-target scores for the hybrid acquisition--selection arms appear in
\cref{tab:coupling_brats_hybrid_centers,tab:coupling_mamamia_hybrid_centers}.

\begin{table}[t]
    \centering
    \caption{\textbf{MAMA-MIA factorial results by seed}, $(B_s,B_t)=(150,10)$. Four-center foreground Dice (\%); matched-seed interaction. Last column: three-seed mean $\pm$ sample std.}
    \label{tab:coupling_mamamia_seeds}
    \footnotesize
    \setlength{\tabcolsep}{3pt}
    \renewcommand{\arraystretch}{1.12}
    \begin{tabular}{@{}lrrrr@{}}
        \toprule
        Target / source & Seed 0 & Seed 1 & Seed 2 & Mean $\pm$ std \\
        \midrule
        Random / Random & 63.48 & 64.20 & 63.13 & \valstd{63.60}{0.55} \\
        TAC / Random & 63.69 & 64.25 & 64.87 & \valstd{64.27}{0.59} \\
        Random / TAC & 63.46 & 64.23 & 65.17 & \valstd{64.29}{0.86} \\
        TAC / TAC & 68.55 & 67.75 & 68.88 & \valstd{68.39}{0.58} \\
        \midrule
        Interaction $\Delta_{\mathrm{int}}$ & +4.88 & +3.47 & +1.96 & \valstd{+3.44}{1.46} \\
        \bottomrule
    \end{tabular}
\end{table}

\subsection{Component Ablations}
\label{app:ablation_protocol}

\begin{table*}[!t]
\centering
\caption{\textbf{BraTS component ablations}, $(B_s,B_t)=(150,10)$. Per-center macro Dice (\%): three-seed mean $\pm$ std; Avg.: unweighted four-center mean.}
\label{tab:ablation_brats_centers}
\footnotesize
\renewcommand{\valstd}[2]{#1\,{\fontsize{6pt}{7pt}\selectfont$\pm$#2}}
\setlength{\tabcolsep}{5pt}
\renewcommand{\arraystretch}{1.10}
\begin{tabular}{@{}lccccc@{}}
\toprule
Variant & C4 & C5 & LGG & GBM & Avg. \\
\midrule
\textbf{Full TAC} & \valstd{79.1}{0.3} & \valstd{78.9}{0.3} & \valstd{67.6}{0.8} & \valstd{83.6}{0.2} & 77.31 \\
\midrule
Image-only curation ($M$ only) & \valstd{75.5}{0.6} & \valstd{77.6}{0.4} & \valstd{61.1}{2.1} & \valstd{83.9}{0.2} & 74.51 \\
Reliability only & \valstd{74.5}{1.0} & \valstd{75.9}{0.3} & \valstd{65.0}{2.5} & \valstd{82.5}{0.3} & 74.47 \\
Coverage only (PRISM) & \valstd{76.9}{0.9} & \valstd{77.1}{0.3} & \valstd{63.8}{1.0} & \valstd{83.2}{0.4} & 75.27 \\
w/o target cap & \valstd{74.6}{0.7} & \valstd{77.9}{0.4} & \valstd{62.0}{0.9} & \valstd{82.4}{0.3} & 74.23 \\
Equal $M/G/P$ weights & \valstd{75.8}{0.8} & \valstd{77.2}{0.5} & \valstd{60.4}{1.8} & \valstd{83.4}{0.3} & 74.20 \\
w/o distribution match ($M$) & \valstd{76.7}{0.9} & \valstd{77.9}{0.3} & \valstd{61.7}{1.9} & \valstd{83.4}{0.2} & 74.91 \\
w/o gradient agreement ($G$) & \valstd{76.8}{0.9} & \valstd{77.5}{0.6} & \valstd{61.2}{2.2} & \valstd{83.4}{0.2} & 74.71 \\
w/o prototype ($P$) & \valstd{76.4}{0.7} & \valstd{77.8}{0.3} & \valstd{60.2}{2.1} & \valstd{83.2}{0.3} & 74.39 \\
\bottomrule
\end{tabular}
\end{table*}

\begin{table*}[!t]
\centering
\caption{\textbf{MAMA-MIA component ablations}, $(B_s,B_t)=(150,10)$. Per-center foreground Dice (\%): three-seed mean $\pm$ std; Avg.: unweighted four-center mean.}
\label{tab:ablation_mamamia_centers}
\footnotesize
\renewcommand{\valstd}[2]{#1\,{\fontsize{6pt}{7pt}\selectfont$\pm$#2}}
\setlength{\tabcolsep}{5pt}
\renewcommand{\arraystretch}{1.10}
\begin{tabular}{@{}lccccc@{}}
\toprule
Variant & NACT & I-SPY1 & Duke & I-SPY2 & Avg. \\
\midrule
\textbf{Full TAC} & \valstd{70.5}{1.4} & \valstd{70.3}{0.4} & \valstd{62.9}{0.8} & \valstd{70.0}{0.5} & 68.39 \\
\midrule
Image-only curation ($M$ only) & \valstd{67.6}{2.2} & \valstd{68.6}{0.4} & \valstd{57.5}{2.0} & \valstd{63.6}{0.7} & 64.32 \\
Reliability only & \valstd{66.7}{2.1} & \valstd{68.1}{0.4} & \valstd{59.3}{2.1} & \valstd{66.1}{0.7} & 65.03 \\
Coverage only (PRISM) & \valstd{65.7}{1.4} & \valstd{70.5}{0.4} & \valstd{58.2}{4.2} & \valstd{62.5}{1.4} & 64.21 \\
w/o target cap & \valstd{67.2}{1.2} & \valstd{69.3}{0.6} & \valstd{58.8}{4.0} & \valstd{67.7}{1.2} & 65.75 \\
Equal $M/G/P$ weights & \valstd{69.8}{1.6} & \valstd{66.9}{0.6} & \valstd{57.3}{3.3} & \valstd{66.5}{1.6} & 65.10 \\
w/o distribution match ($M$) & \valstd{68.1}{1.6} & \valstd{68.1}{0.5} & \valstd{58.4}{2.0} & \valstd{66.5}{1.2} & 65.25 \\
w/o gradient agreement ($G$) & \valstd{70.3}{2.1} & \valstd{71.0}{0.7} & \valstd{58.2}{4.6} & \valstd{67.1}{1.2} & 66.65 \\
w/o prototype ($P$) & \valstd{68.9}{2.2} & \valstd{67.7}{0.4} & \valstd{59.8}{1.6} & \valstd{64.9}{0.6} & 65.31 \\
\bottomrule
\end{tabular}
\end{table*}

\begin{table}[t]
    \centering
    \caption{\textbf{MAMA-MIA hybrid arms by target and seed}, $(B_s,B_t)=(150,10)$, foreground Dice (\%). Per-center scores and four-center seed means.}
    \label{tab:coupling_mamamia_hybrid_centers}
    \scriptsize
    \setlength{\tabcolsep}{2.0pt}
    \renewcommand{\arraystretch}{1.1}
    \begin{tabular}{@{}llrrrrr@{}}
        \toprule
        Target / source & Seed & NACT & I-SPY1 & Duke & I-SPY2 & Avg. \\
        \midrule
        \multirow{3}{*}{TAC / Random} & 0 & 63.79 & 66.15 & 58.84 & 65.98 & 63.69 \\
        & 1 & 64.17 & 66.93 & 59.60 & 66.31 & 64.25 \\
        & 2 & 65.07 & 67.14 & 60.30 & 66.97 & 64.87 \\
        \midrule
        \multirow{3}{*}{Random / TAC} & 0 & 66.56 & 65.57 & 57.35 & 64.35 & 63.46 \\
        & 1 & 65.34 & 66.64 & 59.65 & 65.28 & 64.23 \\
        & 2 & 66.20 & 67.62 & 61.07 & 65.80 & 65.17 \\
        \midrule
        \multirow{3}{*}{TAC / CoreSet} & 0 & 68.82 & 66.95 & 52.74 & 65.95 & 63.62 \\
        & 1 & 66.25 & 67.53 & 55.86 & 67.13 & 64.19 \\
        & 2 & 67.21 & 67.12 & 54.64 & 67.68 & 64.16 \\
        \midrule
        \multirow{3}{*}{CoreSet / TAC} & 0 & 69.76 & 65.21 & 52.03 & 66.84 & 63.46 \\
        & 1 & 68.64 & 65.56 & 60.26 & 64.18 & 64.66 \\
        & 2 & 69.38 & 65.42 & 60.04 & 64.46 & 64.83 \\
        \midrule
        \multirow{3}{*}{TAC / ORIENT} & 0 & 66.04 & 70.56 & 58.50 & 64.57 & 64.92 \\
        & 1 & 66.56 & 69.45 & 57.60 & 64.68 & 64.57 \\
        & 2 & 67.06 & 70.05 & 58.41 & 62.08 & 64.40 \\
        \midrule
        \multirow{3}{*}{CLUE / TAC} & 0 & 69.93 & 68.41 & 56.76 & 66.93 & 65.51 \\
        & 1 & 67.29 & 68.25 & 58.09 & 66.99 & 65.16 \\
        & 2 & 67.29 & 68.53 & 58.61 & 65.80 & 65.06 \\
        \bottomrule
    \end{tabular}
\end{table}

Component ablations hold the TAC-acquired target support, frozen
selection-time evidence, split manifests, final segmentation protocol, and
checkpoint-selection rule fixed. The controls in
Table~4 of the main paper are:
\begin{itemize}
\item \textbf{Image-only curation:} set $R_j=M_j$ in $F$, retaining
the support-capped similarities and all acquired target annotations
for final training.
\item \textbf{Reliability only:} select the top $B_s$ cases by $R_j$.
\item \textbf{Coverage only (PRISM):} set $R_j=1$, yielding the PRISM-FLVMI objective with $\eta=1$ under the same target support, similarity kernel, source budget, and training protocol.
\item \textbf{Without the target cap:} replace $h_{ij}$ by $K_{ij}$
inside $F$, retaining target-conditioned reliability.
\item \textbf{Equal-weight fusion:} use
$\lambda_M=\lambda_G=\lambda_P=1/3$ without renormalizing the fused score.
\item \textbf{Leave-one-signal-out:} remove $M$, $G$, or $P$ and
renormalize the remaining weights to sum to one while preserving
their relative proportions.
\end{itemize}
All evaluated controls use the same three independent training seeds.
Tables~\ref{tab:ablation_brats_centers} and~\ref{tab:ablation_mamamia_centers}
report the per-center results; their Avg.\ means are unweighted averages of
the four center means. The main-paper averages and differences from full TAC
are rounded to one decimal place.

\section{Limitations and Existing Assets}
\label{app:limitations_impact}
Evaluation covers four held-out centers per benchmark on BraTS 2021 and
MAMA-MIA, spanning different anatomies, labels, and final segmenters.
Generalization to other centers and imaging modalities remains an empirical
question. Selection depends on the full-source warm-up, acquired annotations,
and frozen features and gradients. The fused reliability score is used as a relative utility weight rather
than a calibrated probability of transfer;
the greedy guarantee applies to the fixed selection objective. The three-seed comparisons characterize pipeline variability on fixed patient
splits, not uncertainty from test-patient or center sampling or from annotation
quality. Clinical deployment
requires separate validation, human oversight, and privacy-preserving data governance.

\subsection{Existing Assets and Data Access}
\label{app:assets}
We use BraTS 2021~\citep{baid2021brats2021} and
MAMA-MIA~\citep{garrucho2025mamamia} under their data-use conditions,
with PyTorch~\citep{paszke2019pytorch} implementations.
Restricted clinical data and third-party pretrained weights are not
redistributed and must be obtained from official sources under the
applicable licenses and access terms.

\subsection{Data Ethics and Use Conditions}
\label{app:data_ethics}
This study uses previously released, de-identified BraTS 2021 MRI data
\citep{baid2021brats2021,menze2015multimodal,bakas2017advancing,baid2023bratsdataset}
and MAMA-MIA data
\citep{garrucho2025mamamia,newitt2016ispy1,newitt2016nact,li2022ispy2,saha2021duke}.
No participants were recruited and no new patient data were collected.
The BraTS 2021 challenge release is distributed under CC BY 4.0;
the MAMA-MIA benchmark uses CC BY-NC 4.0, reflecting the most
restrictive of its four source collections.
We respect the original attribution and access conditions, do not
attempt patient re-identification, and do not redistribute restricted
source images. Ethical oversight and consent for the source cohorts
are documented by the data providers.

\clearpage
\endgroup

\end{document}